\documentclass[aps,prl,nobibnotes,twocolumn,superscriptaddress,bibliography]{revtex4-2}

\usepackage{amsfonts}
\usepackage{mathrsfs}
\usepackage{amsmath}
\usepackage{color}
\usepackage{graphicx}
\usepackage{bm}
\usepackage{amssymb}
\usepackage{xspace}
\usepackage{epstopdf}
\usepackage{dcolumn}
\usepackage{longtable}
\usepackage{multirow}
\usepackage{float}
\usepackage{comment}

\usepackage[colorlinks=true, letterpaper=true, pdfstartview=FitV, linkcolor=blue, citecolor=blue, urlcolor=blue]{hyperref}

\makeatother

\begin{document}

\title{Cornertronics in Two-Dimensional Second-Order Topological Insulators}


\author{Yilin Han}
\thanks{These authors contributed equally to this work.}
\affiliation{Centre for Quantum Physics, Key Laboratory of Advanced Optoelectronic Quantum Architecture and Measurement (MOE), School of Physics, Beijing
Institute of Technology, Beijing 100081, China}
\affiliation{Beijing Key Lab of Nanophotonics \& Ultrafine Optoelectronic Systems, School of Physics, Beijing Institute of Technology, Beijing 100081, China}

\author{Chaoxi Cui}
\thanks{These authors contributed equally to this work.}
\affiliation{Centre for Quantum Physics, Key Laboratory of Advanced Optoelectronic Quantum Architecture and Measurement (MOE), School of Physics, Beijing
Institute of Technology, Beijing 100081, China}
\affiliation{Beijing Key Lab of Nanophotonics \& Ultrafine Optoelectronic Systems, School of Physics, Beijing Institute of Technology, Beijing 100081, China}

\author{Xiao-Ping Li}
\affiliation{School of Physical Science and Technology, Inner Mongolia University, Hohhot 010021, China}

\author{Ting-Ting Zhang}
\affiliation{Beijing National Laboratory for Condensed Matter Physics, Institute of Physics, Chinese Academy of Sciences, Beijing 100190, China }

\author{Zeying Zhang}
\affiliation{College of Mathematics and Physics, Beijing University of Chemical Technology, Beijing 100029, China}

\author{Zhi-Ming Yu}
\email{zhiming\_yu@bit.edu.cn}
\affiliation{Centre for Quantum Physics, Key Laboratory of Advanced Optoelectronic Quantum Architecture and Measurement (MOE), School of Physics, Beijing
Institute of Technology, Beijing 100081, China}
\affiliation{Beijing Key Lab of Nanophotonics \& Ultrafine Optoelectronic Systems, School of Physics, Beijing Institute of Technology, Beijing 100081, China}

\author{Yugui Yao}
\email{ygyao@bit.edu.cn}
\affiliation{Centre for Quantum Physics, Key Laboratory of Advanced Optoelectronic Quantum Architecture and Measurement (MOE), School of Physics, Beijing
Institute of Technology, Beijing 100081, China}
\affiliation{Beijing Key Lab of Nanophotonics \& Ultrafine Optoelectronic Systems, School of Physics, Beijing Institute of Technology, Beijing 100081, China}

\begin{abstract}
Traditional electronic devices rely on the electron's intrinsic degrees of freedom (d.o.f.)  to process information. However, additional d.o.f. like the valley,  can emerge in the low-energy states of certain systems.
Here, we show that the quantum dots (QDs) constructed from two-dimensional (2D) second-order topological insulators (SOTI) posses a new  kind of  d.o.f., namely corner freedom,  related  to the topological corner states that reside at  different corners of the systems.
Since the corner states are well separated in real space, they can be  individually and intuitively manipulated, giving rise to the concept of cornertronics.
Via symmetry analysis and material search, we identify the TiSiCO-family monolayers as the first prototype of cornertronics materials, where the  corner states can be controlled by both electric and optical fields, due to  novel corner-layer coupling (CLC)  effect and  corner-contrasted linear dichroism.
Furthermore, we find  that the  band gap of the  TiSiCO nanodisk lies in the terahertz region and is robust to size reduction.
These results indicate that the TiSiCO nanodisks can be used to design terahertz devices with ultrasmall size and electric-field tunable band gap.
Besides, the TiSiCO nanodisks are  simultaneously sensitive to both the strength and  polarization of the terahertz waves.
Our findings not only pave the way for cornertronics, but also open a new direction for research in 2D SOTI,  QD and terahertz electronics.
\end{abstract}
\maketitle

{\emph{\textcolor{blue}{Introduction.--}}}
Degrees of freedom is a fundamental concept in physics referring to the independent parameters of  systems \cite{neil1976solid}.
Generally, each d.o.f. has its own unique physical properties and enables the design of electronic devices with novel functions \cite{ZwanenburgRMP-2013,zutic2004spintronics,baltz2018antiferromagnetic,schaibley2016valleytronics}.
Hence, it is always fascinating to reveal new kinds of freedom in realistic physical systems.

The traditional devices use only electronic charge to process  information \cite{joachim2000electronics,wu2007graphenes,wang2012electronics,fiori2014electronics}.
By introducing spin d.o.f., the functionality of the electronic devices was  greatly expanded \cite{bader2010spintronics,wolf2001spintronics,chappert2007emergence,awschalom2007challenges}.
Later, it  was found that the electrons  may exhibit emergent d.o.f. in the low-energy of certain systems, such as valley, which appears in the systems with multiple energy  extrema in the momentum space \cite{rycerz2007valley,xu2014spin,sui2015gatetunable,tong2016concepts,yu2020valleylayer}.
Similar to spin, the valley leads to  many unique phenomena that can not be realized in other systems \cite{xiao2012coupleda,zhu2012fieldinduced,jiang2013generation,
grujic2014spinvalley,qiao2014current,pan2015valleypolarized,pan2015perfect,settnes2016graphene,cheng2018manipulation,li2018valley}.
The study of spin and valley d.o.f. has received extensive attention, giving rise  to the emergence of spintronics and valleytronics, respectively \cite{bader2010spintronics,rycerz2007valley}.

Recently, a new class of topological insulators--second-order topological insulators (SOTI)--has been  proposed \cite{benalcazar2017quantized,langbehn2017reflectionsymmetrica,song2017ensuremath2,
wang2018hightemperature,schindler2018higherordera,ezawa2018magnetic,ezawa2018topological,wu2022quantized}, in which the topology of $d$-dimensional bulk states manifests on $(d-2)$-dimensional boundaries.
Hence, for a 2D SOTI, the topologically protected boundary states will appear at the intersection of two crystal edges, termed as corner states \cite{benalcazar2017quantized}.
Since the discovery of graphdiyne  \cite{sheng2019twodimensional,RN2830}, many  materials have been predicted as 2D SOTI, including both non-magnetic \cite{guo2022quadrupole,reis2017bismuthene,
chen2021graphynea,liu2019twodimensionalb,park2019higherordera,xue2021higherordera,takahashi2021general,
qian2022c,zeng2021multiorbitala,li2022secondorder,qian2021secondorderb,li2023robusta,costa2021discovery,liu2022secondorder}
and magnetic materials \cite{chen2020universala,ren2020engineering,RN2831,mao2023ferroelectric,zhang2023magnetic,adma.202402232}.
The corner states also can be realized in second-order topological superconductors \cite{pahomi2020braiding,zhang2020topological,li2022higherorder}, where they can be manipulated by their spatial distribution.

\begin{figure}[t]
\includegraphics[width=0.48\textwidth]{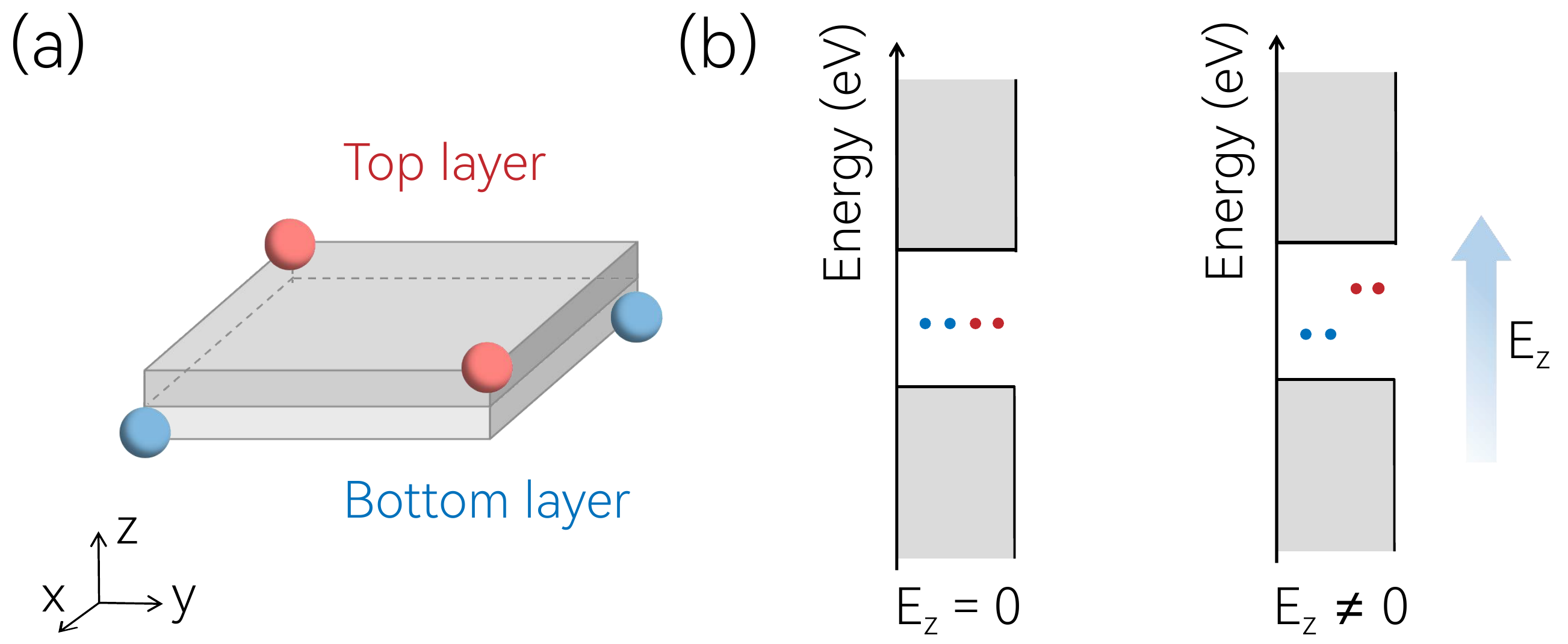}
\caption{\label{fig: illu} (a) Schematic showing a 2D SOTI nanodisk with CLC effect, where the corner states have finite but opposite layer polarization $P_z$.
The corner state with $P_z>0$ ($P_z<0$) is  mainly localized on the top (bottom) layer. (b) Spectrum of  a 2D SOTI nanodisk with topological corner states occurring  in the band gap.  The shadow regions denote the energy levels out of  the band gap. The red (blue) dots denote the degenerate corner states with  $P_z>0$ ($P_z<0$).
Under a gate electrical field  $E_z$, the  corner states with opposite $P_z$ will have different energy shifts, leading to electric field control of  corner polarization.}
\end{figure}

In this work, we show that the topological  corner states  can be considered the real-space counterpart of the valley states, as they are low-energy states that reside in bulk band gap and   at different corners of the SOTI nanodisk, well separated  in real space (see Fig. \ref{fig: illu}).
This also means that the corner state is quite stable, protected by the bulk band gap and their spatial separation.
Besides, each corner can be individually and intuitively manipulated by local external fields.
These strongly suggest that the low-energy electrons of the SOTI nanodisks are endowed with a new and stable  d.o.f., namely,  the corner degree.

To enable the corner d.o.f., two  prerequisites should be satisfied, namely, revealing unique  phenomena associated with the corner states and finding  efficient ways  to generate corner polarization.
However, both  have not been well investigated in previous works.
For the former, since the SOTI nanodisk is a  kind of QD, it is instructive  to discover signatures distinct from the conventional QDs \cite{ameta2023quantum,jung2021prospectsb,medintz2005medintz,yinglim2015carbon,bera2010quantum,eich2018spin,kurzmann2019charge,banszerus2020electron,jung2021prospectsb,moreels2009sizedependent,D0NR06523D,rider2021proposal}.
Regarding the latter, we need an  external field, to which the corner states have opposite responses.
Amongst all  possible external fields,  the gate electric field ($E_z$) normal to the SOTI QD is probably the most ideal, due to  its convenience and compatibility.
Hence, it is important to  identify suitable SOTI materials with  two sets of corners that have the opposite responses to  $E_z$.
This is not a trivial task.
While the material prediction of 2D SOTIs has been intensively studied,  the previously reported materials can not satisfy this task.

Here, we demonstrate that the corner states in the SOTI with novel CLC can have an opposite response to $E_z$. We  present  general symmetry requirements for the CLC, and identify  the TiSiCO-family monolayers as the first prototype of  the 2D SOTI with CLC effect.
Specifically, we find that the  TiSiCO-family monolayers are  2D SOTIs characterized by nontrivial real Chern number and multiple four-fold degenerate   corner states.
Particularly, the four degenerate corner states, connected by  $S_{4z}$ symmetry, have strong layer polarization and are mainly localized at the  top or bottom layer, as illustrated in Fig. \ref{fig: illu}(a).
Thus, the degenerate corner states can be  divided into two groups with different  layer polarization.
By  applying a gate field $E_z$, the two groups  will feature opposite energy shifts, leading to a static, continuous, switchable and  wide-range control of corner polarization.

We also show that  the monolayer Ti$_2$SiCO$_2$ (ML-TSCO)  QD  has many other unique physical properties that are absent in the  conventional nano-devices and QDs, including corner-contrasted optical selection rules and robust small band gap against size reduction.
For example, when the length of the square ML-TSCO nanodisk is about $2$ nm,
the energy splitting of the corner states  is still as small as $\sim1$ meV ($0.25$ THz).
All these results suggest that the ML-TSCO nanodisks can be used to design novel ultrasmall terahertz sensors with electric field tunable band gap and simultaneous sensitivity to the field strength and polarization of the terahertz waves.

{\emph{\textcolor{blue}{General analysis of CLC.--}}}
We consider a 2D SOTI nanodisk with a finite thickness  that lies in the $x$-$y$ plane.
The layer polarization of an  electronic state $|\psi\rangle$ can be defined as \cite{yu2020valleylayer}
\begin{eqnarray}
P_z& = & \langle \psi|{\cal{P}}|\psi\rangle,
\end{eqnarray}
where  ${\cal{P}}$ is a diagonal matrix representing layer operation.
${\cal{P}}_{ii}=1$ if the  $i$-th basis of $|\psi\rangle$ is above  the middle plane of the system  and ${\cal{P}}_{ii}=-1$ if it is below the middle plane.
$P_z>0$ ($P_z<0$ ) indicates that $|\psi\rangle$ distributes more weight in the top (bottom) layers [see Fig. \ref{fig: illu}(a)].

Due to the higher-order topology, a SOTI nanodisk  exhibits  multiple topological corner states.
If these  corner states can be divided into two groups with opposite $P_z$, the corner and layer d.o.f. are  coupled, leading to the CLC effect.
Obviously, under a gate field $E_z$, the corner states with CLC will have an opposite energy shift, as shown in Fig. \ref{fig: illu}(b). This corner-contrasted electric response enables the gate field control of corner polarization.
To achieve CLC, the following conditions should be satisfied.

(i) The layer polarization of the corner states must be finite. Hence, the 2D materials should not have  horizontal mirror $M_z$, which excludes all the 2D SOTIs having only one atomic layer.

(ii) For simplicity, we assume there are only two or two groups of corner states with opposite $P_z$, denoted as $C_u$ and $C_d$, respectively. The  symmetry operations of the nanodisk  can be divided into two classes: changing $z$ to $-z$ (${\cal{R}}_-$) and keeping $z$ invariant (${\cal{R}}_+$).
Since $P_z$ is invariant (switched) under ${\cal{R}}_+$ (${\cal{R}}_-$), the system should have
\begin{eqnarray}
{\cal{R}}_{+}C_{u(d)}=C_{u(d)}, &\ \ \ & {\cal{R}}_{-}C_{u(d)}=C_{d(u)} \label{eq:two}.
\end{eqnarray}

According to the above symmetry discussion, we  go through all the 31 layer point groups (LPGs), which describe the point-group symmetry of the 2D system with finite thickness \cite{zhang2023encyclopedia}.
 We find that 11 LPGs can  host the novel CLC effect, which are $C_{2,\parallel}$, $D_{n}$ ($n=2,3,4,6$), $C_{2h}$, $S_{n}$ ($n=2,4,6$), $D_{2d}$ and $D_{3d}$.
The possible configurations of the corner states in these 11 LPGs are illustrated in Supplemental Materials (SM) \cite{hansupplemental}.

{\emph{\textcolor{blue}{Material realization.--}}}
We then demonstrate that the   TiSiCO-family monolayer $X_2Y$CO$_2$ ($X$=Ti, Zr, Hf; $Y$=Si, Ge) are the SOTI with novel CLC.
Since all the monolayers have similar crystalline structure and topological properties, we use ML-TSCO as an example to show the CLC effect.
The ML-TSCO was predicted as  a valleytronics material \cite{yu2020valleylayer},
however, its topological properties have  not been studied.

\begin{figure}[t]
\includegraphics[width=0.48\textwidth]{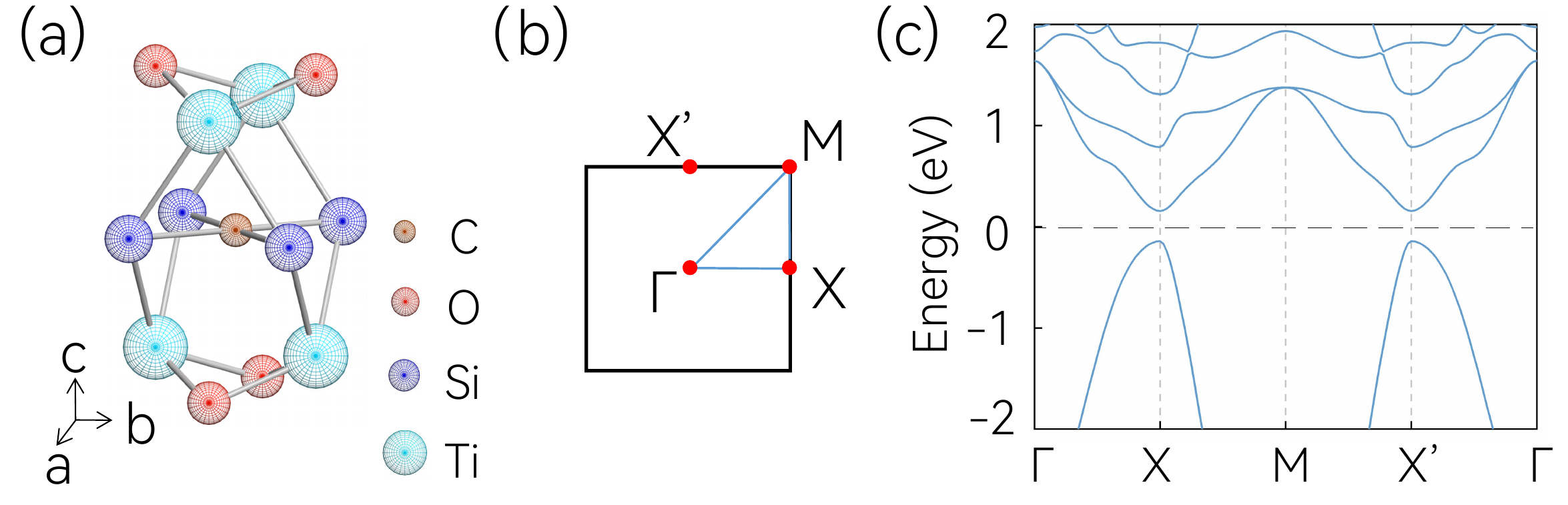}
\caption{\label{fig: TiSiCO} (a) Crystal structure of the ML-TSCO. (b) The corresponding 2D BZ. (c) Band structure of the ML-TSCO.}
\end{figure}

The crystalline structure of ML-TSCO is shown in Fig. \ref{fig: TiSiCO}(a), having five atomic layers with a height of $\sim4$ \AA.
Its lattice constant is calculated as $a=b=2.82$ \AA \ \cite{hansupplemental}.
Particularly, the ML-TSCO has time-reversal symmetry (${\cal{T}}$) and belongs to the $D_{2d}$ point group, which is one of the 11 target LPGs.
Due to negligible spin-orbit coupling (SOC), the SOC  effect in the ML-TSCO can be neglected.
The  calculated band structure of the ML-TSCO without SOC is plotted in Fig. \ref{fig: TiSiCO}(c), from which one observes that the ML-TSCO is a semiconductor with two valleys located at $X$ and $X'$ points.

Because  the ML-TSCO perseveres both $C_{2z}$ and  $\mathcal{T}$, its topology can be characterized by the real Chern number $\nu_R$, which is a $\mathbb{Z}_2$ invariant \cite{ahn2018band,zhaoSymmetricRealDirac2017}.
Both the parity criterion and the Wilson loop method are used to verify the nontrivial $\nu_R$.
For the parity criterion, we analyze the $C_{2z}$ eigenvalues of the electronic bands at the four $C_{2z}$-invariant momentum points $\Gamma_i$ ($i=1,2,3,4$), which are $\Gamma$, M, $X$ and $X’$ in the Brillouin zone (BZ).
Then, $\nu_R$ is obtained as \cite{ahn2018band,zhuPhononicRealChern2022a}
\begin{eqnarray}
(-1)^{\nu_R}=\prod_{i=1}^4(-1)^{\lfloor N_{\mathrm{occ}}^{-}\left(\Gamma_i\right) / 2\rfloor},\label{eq:four}
\end{eqnarray}
where  $\lfloor ... \rfloor$ is the floor function and $N_{\mathrm{occ}}^{-}$ denotes the number of the occupied  bands with   $C_{2z}=-1$ at $\Gamma_i$  point.
For the ML-TSCO, $N_{\mathrm{occ}}^{-}$ are 10, 11, 11 and 8 for $\Gamma$, $X$, $X’$ and $M$ points, respectively. Therefore,   $\nu_R=1$ is obtained, indicating that ML-TSCO is a  real Chern insulator.

\begin{figure}[t]
\includegraphics[width=0.48\textwidth]{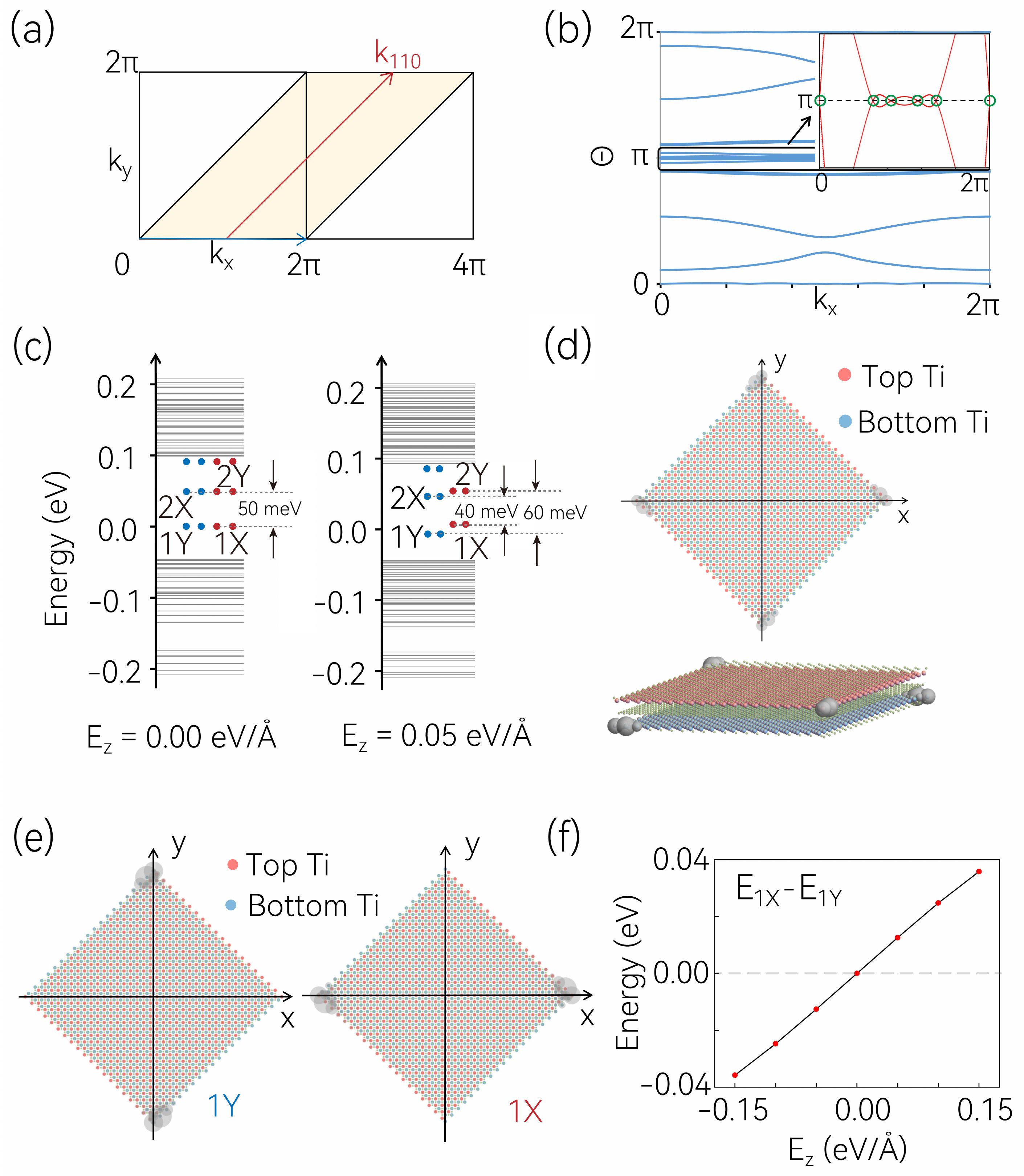}
\caption{\label{fig:corner state} (a) The Wilson loop is calculated along an orientable circle (the red path). (b) Calculated Wilson loop for ML-TSCO.  (c) Energy spectrum of ML-TSCO nanodisk with $S_4$ symmetry for $E_z=0$ and $E_z=0.05$ eV/\AA.
The solid lines  denote the energy levels out of the band gaps of both bulk and edge states.
Red and blue dots denote the corner states mainly localized at the top and bottom Ti atoms, respectively.
(d) Charge density distribution  of the lowest corner states in real space for $E_z=0$. (e) Charge density distribution in real space of the $1X$  and $1Y$ corner states for $E_z=0.05$ eV/\AA. The real-space distributions of $2X$  and $2Y$ states are shown in SM \cite{hansupplemental}.
(f) The energy difference between $1X$  and $1Y$ corner states versus  $E_z$.}
\end{figure}

The real Chern number  $\nu_R$ also can  be obtained by the Wilson-loop method \cite{ahn2018band}.
However, we find that  the occupied bands  of the ML-TSCO are non-orientable along both $k_x$ and $k_y$ directions, due to the nontrivial $\pi$ Zak phase   \cite{hansupplemental}.
Alternatively, we calculate the Wilson loop along $(110)$ direction [see Fig. \ref{fig:corner state}(a)], which is  an orientable circle \cite{ahn2018band}.
The calculated Wilson loop is shown in Fig. \ref{fig:corner state}(b), where odd (five) crossing points at $\theta = \pi$ axis can be observed.
This means $\nu_R=1$ and again proves that the ML-TSCO is a 2D SOTI.

To study the topological corner states and the CLC effect, we construct a Wannier tight-binding (TB) model of the ML-TSCO based on the first-principles calculations.
In Fig. \ref{fig:corner state}(c), the spectrum of a square nanodisk with $S_{4z}$ symmetry is presented.
One observes that inside the  band gap of bulk and edge states, there are three fourfold degenerate states.
We have checked that  all these degenerate  states are  corner states.
Here, we term the corner at  $x$-axis ($y$-axis) $X$ ($Y$) corner.
Since the four corners of the  ML-TSCO  has only ${\cal{T}}$ symmetry, and are connected by $S_{4z}$, which is a symmetry in $R_{-}$, the $X$ and $Y$ corner states must have finite but opposite layer polarizations.
The spatial distribution of the lowest degenerate states [marked as $1X$ and $1Y$ in Fig. \ref{fig:corner state}(c)]  are plotted in Fig. \ref{fig:corner state}(d).
One observes that the two $1X$ corner states  at the $x$-axis  are mainly distributed in the top Ti atoms ($P_z>0$), while the two $1Y$ corner states at the $y$-axis have more distribution on the bottom Ti atoms ($P_z<0$), consistent with the above analysis.
This result undoubtedly confirms the existence of CLC.
Interestingly, $2X$ ($2Y$) states are mainly distributed in the bottom (top) layer \cite{hansupplemental}.
The topological corner states generally are robust against perturbations \cite{sheng2019twodimensional,hansupplemental} and have been experimentally observed in metal-organic frameworks \cite{hu2023identifying}.

{\emph{\textcolor{blue}{Corner polarization by gate field.--}}}
We then use the Wannier TB to demonstrate that the CLC can generate corner polarization by gate electric field.
Here, the gate field is approximately represented as an on-site energy $\Delta=\alpha E_z r_z$ in the Wannier TB.  $r_z$ is the $z$-component of each atomic position, and $\alpha \approx 0.07$ \ is obtained by comparing the  bulk band structures obtained from first-principles calculations and Wannier TB when a small $E_z$ is applied.
The $S_{4z}$ is broken by $E$-field but the $C_{2z}$ is maintained.
As a result, the four degenerated corner states are split into two doubly degenerate modes.
For $E_z=0.05$ eV/\AA~(achievable in current experiments \cite{zhang2009direct,bampoulis2023quantum}), the   $1Y$ ($2X$) corner states  are pulled down while the  $1X$  ($2Y$) corner states  are pushed up, as  shown in Fig. \ref{fig:corner state}(c,f).
This means that the $E$-field is an efficient method to control the energy splitting between  the $1X$  and $1Y$ corners, as well as the intra-corner band gap between $1X$ ($1Y$) and $2X$ ($2Y$).
With suitable doping, the low-energy states of the nanodisk will only occupy the  $Y$ corner, leading to a corner polarization.
By  reversing  $E$-field, only the $X$ corner is  occupied and the corner polarization is switched.

\begin{figure}[t]
\includegraphics[width=0.48\textwidth]{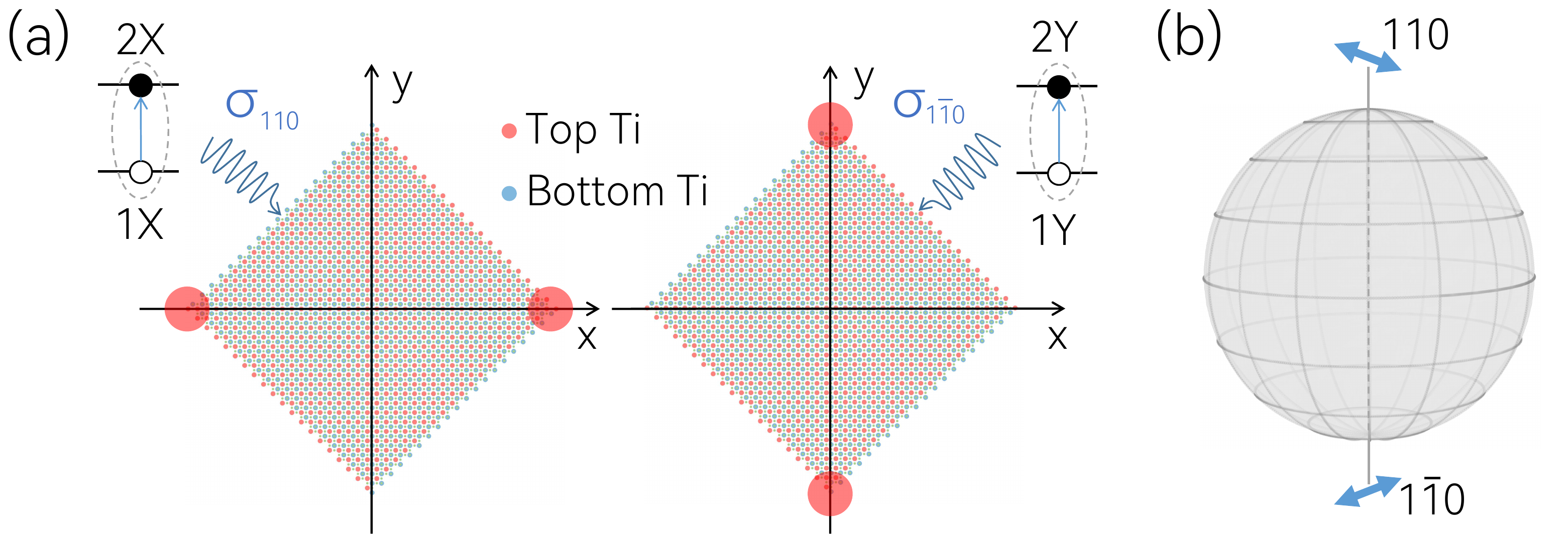}
\caption{\label{fig:optic} (a) Corner-contrasted optical selection rule of ML-TSCO. (b) Corresponding pseudospin (corner) states plotted on the Bloch sphere. The exciton at the $X$ ($Y$) corner corresponds to pseudospin up (down), which is excited by $\sigma_{110}$ ($\sigma_{1\bar{1}0}$) linearly polarized light.}
\end{figure}

{\emph{\textcolor{blue}{Corner-contrasted optical selection rule.--}}} Here, we consider the intra-corner optical transitions from $1X$ ($1Y$) to $2X$ ($2Y$) [see Fig. \ref{fig:corner state}(c)].
Remarkably, the intra-corner band gap  is about $50$ meV ($\sim$12 THz),  indicating that  the ML-TSCO nanodisk can be used to detect and emit  terahertz waves.

In the dipole approximation, the strength of the optical interband absorption is determined by \cite{perinetti2011optical}
\begin{eqnarray}\label{eq:Op}
{\bm M}^{X(Y)} &\propto& \langle \psi_{2X(2Y)}|{\hat{\bm{r}}}|\psi_{1X(1Y)}\rangle,
\end{eqnarray}
where  $\hat{\bm{r}}$ is the position operator.
Interestingly, we approximately have   the linear polarization
\begin{eqnarray}
\eta_{110}&\equiv& \frac{|M_{110}^X|^2-|M_{1\bar10}^X|^2}{|M_{110}^X|^2+|M_{1\bar10}^X|^2}\simeq 1,
\end{eqnarray}
for the $X$  corner.
This phenomenon can be quantitatively understood as the two states $\psi_{2X}^*$ and $\psi_{1X}$  forming an electric dipole, with dipole moment predominantly oriented along $(110)$ direction.
Since the ML-TSCO nanodisk has $S_{4z}$, we must have
\begin{eqnarray}
\eta_{1\bar10}&\equiv& \frac{|M_{1\bar10}^Y|^2-|M_{110}^Y|^2}{|M_{1\bar10}^Y|^2+|M_{110}^Y|^2}\simeq 1,
\end{eqnarray}
for the $Y$  corner [see Fig. \ref{fig:optic}].
Therefore, the interband transitions  at   $X$ and $Y$  corners are respectively coupled exclusively with ($110$)-linearly ($\sigma_{110}$) and ($1\bar10$)-linearly ($\sigma_{1\bar10}$) polarized light, leading to corner-contrasted linear dichroism.
For the ML-TSCO nanodisk, one can selectively excite the carriers in a certain corner  by controlling the polarization of light.
Besides, since the optical responses generally are proportional to the field strength, the ML-TSCO nanodisk may be the first QD candidate that is   sensitive to both  strength and polarization of terahertz waves.
In contrast, to achieve this function  in  conventional QDs, auxiliary equipment  is needed \cite{shi2022roomtemperature}.

\begin{figure}[t]
\includegraphics[width=0.48\textwidth]{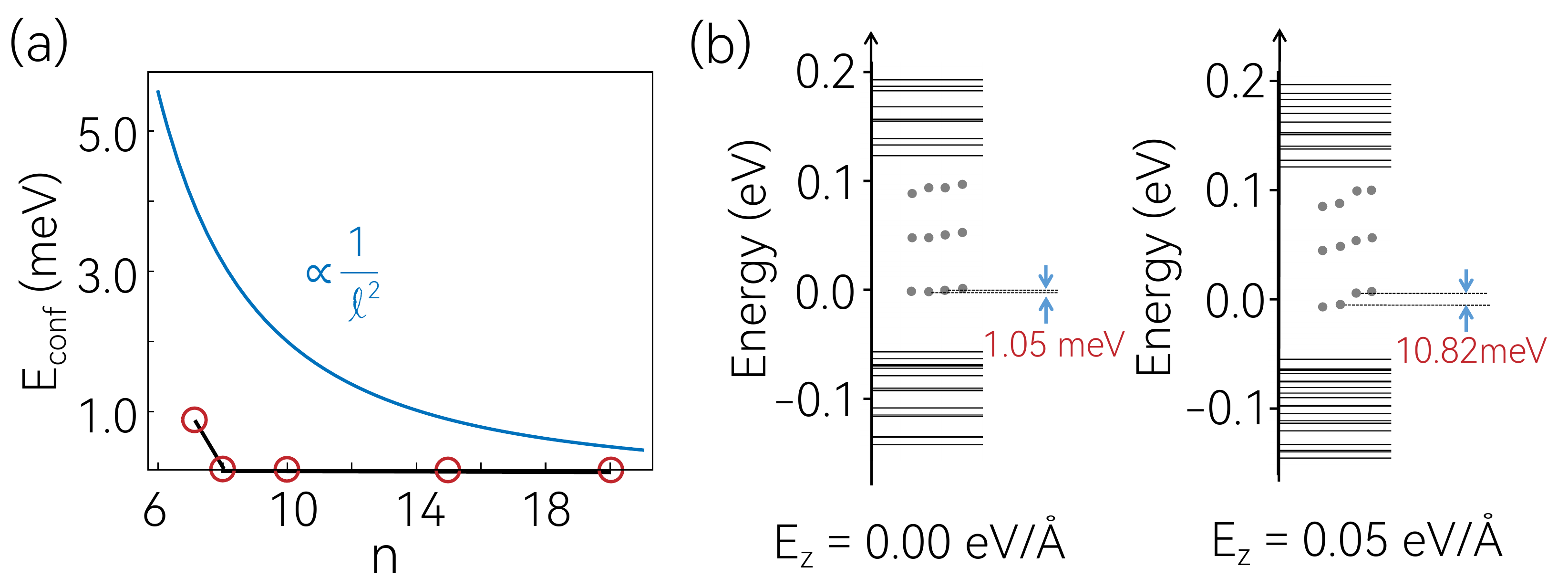}
\caption{\label{fig:gap} (a) The confinement (splitting) energy $E_{\text{conf}}$ (red circles) of the lowest corner states for the  ML-TSCO nanodisk with a length $l=(n-1/2)\times\sqrt{2}a$ , which is exponentially suppressed with the nanodisk size.
In contrast, $E_{\text{conf}}\propto 1/l^2$ for  conventional QD, as illustrated by the blue curve.
(b) The energy spectrum of the  ML-TSCO nanodisk with $n=7$ for $E_z= 0 $ and $E_z= 0.05$  eV/\AA.
Here, the corner states are strongly hybridized due to the small size, in such case, the corner d.o.f. can not be well defined.}
\end{figure}

{\emph{\textcolor{blue}{Robust small band gap in SOTI-based QDs.--}}}
The band gap or confinement energy is an important characteristic parameter of the QDs \cite{medintz2005medintz,yinglim2015carbon,bera2010quantum,eich2018spin,kurzmann2019charge,banszerus2020electron}.
Approximately, the confinement energy $E_{\text{conf}}$, defined as the difference between  QD  and  bulk band gaps,  is inversely proportional to the square of  the QD size \cite{jung2021prospectsb}, as illustrated in Fig. \ref{fig:gap}(a).
Therefore, an ultrasmall QD is not likely to have a band gap in the terahertz range. Here, we show that the SOTI-based QDs can fundamentally solve this challenge.

For sufficiently large ML-TSCO nanodisks, all the corner states are  four-fold degenerate, and the system  is a metal when the  corner states at the Fermi level are half occupied.
For clarity,  we assume the lowest corner states ($1X$ and $1Y$) are half-occupied [see Fig. \ref{fig:corner state}(c)].
By reducing the  size of the nanodisk, the degenerate corner states will be split due to  quantum confinement, and the system becomes an insulator with a global  band gap $E_g=E_{\text{conf}}$.
However, since the  corner states are  localized at different corners of the nanodisk and  are well separated from each other, the  global  band gap will be   exponentially suppressed with the nanodisk size.
Similarly, the  intra-corner band gap is also not sensitive to the nanodisk size.
This is  completely different from the  conventional QDs.

We also calculate the  spectrum of the ML-TSCO nanodisk with the same square  geometry but different length, and find that the lowest corner states are still almost degenerate for the nanodisk with $n=10$ \cite{hansupplemental}.
For the  nanodisk  with $n=7$ ($l\simeq2.59$ nm), we have $E_{\text{g}}=1.05$ meV [see Fig. \ref{fig:gap}(b)].
In comparison,  the band gap of PbS QD with a radius of about 3 nm exceeds 1.2 eV \cite{jung2021prospectsb,moreels2009sizedependent}.
Meanwhile, the intra-corner band gap is almost unchanged when  $l$ decreases \cite{hansupplemental}.
All these results indicate that the ML-TSCO nanodisk with ultrasmall size can still have a terahertz band gap, and would  have great advantages in designing ultrasmall terahertz devices.
We can also integrate many ML-TSCO nanodisks into a single device to increase sensitivity without making the device bulky.

Moreover,  both global and intra-corner  band gaps of the ML-TSCO can be tuned by the gate field. For  $E_z=0.05$ eV/\AA, the global band gap  of the nanodisk  with  $n=7$ nm reaches  $10.82$ meV [see Fig. \ref{fig:gap}(b)], and the intra-corner band gap of the nanodisk  with  large $n$ changes by $\sim 10$ meV [see Fig. \ref{fig:corner state}(c)].
This means that  the emission and absorption wavelength of the ML-TSCO nanodisk   can be further controlled by the gate field.
This is another advantage  of the  ML-TSCO nanodisk.

{\emph{\textcolor{blue}{Conclusion.--}}}
In this work, we propose  corner d.o.f. in the low-energy states of the 2D SOTI nanodisks,  and present the CLC effect  to realize a static control of the  corner polarization by gate electric field.
Based on first-principle calculations, we predict the  TiSiCO-family monolayers as the first example of the 2D SOTI with CLC.
We also find that the ML-TSCO nanodisk  exhibits corner-contrasted linear dichroism and unusual  confinement energy behavior under size reduction.
Our results suggest that  the TiSiCO-family monolayers and other SOTIs with CLC are ideal materials for  developing novel terahertz devices.


\bibliography{mybib}

\begin{thebibliography}{95}%
\makeatletter
\providecommand \@ifxundefined [1]{%
 \@ifx{#1\undefined}
}%
\providecommand \@ifnum [1]{%
 \ifnum #1\expandafter \@firstoftwo
 \else \expandafter \@secondoftwo
 \fi
}%
\providecommand \@ifx [1]{%
 \ifx #1\expandafter \@firstoftwo
 \else \expandafter \@secondoftwo
 \fi
}%
\providecommand \natexlab [1]{#1}%
\providecommand \enquote  [1]{``#1''}%
\providecommand \bibnamefont  [1]{#1}%
\providecommand \bibfnamefont [1]{#1}%
\providecommand \citenamefont [1]{#1}%
\providecommand \href@noop [0]{\@secondoftwo}%
\providecommand \href [0]{\begingroup \@sanitize@url \@href}%
\providecommand \@href[1]{\@@startlink{#1}\@@href}%
\providecommand \@@href[1]{\endgroup#1\@@endlink}%
\providecommand \@sanitize@url [0]{\catcode `\\12\catcode `\$12\catcode
  `\&12\catcode `\#12\catcode `\^12\catcode `\_12\catcode `\%12\relax}%
\providecommand \@@startlink[1]{}%
\providecommand \@@endlink[0]{}%
\providecommand \url  [0]{\begingroup\@sanitize@url \@url }%
\providecommand \@url [1]{\endgroup\@href {#1}{\urlprefix }}%
\providecommand \urlprefix  [0]{URL }%
\providecommand \Eprint [0]{\href }%
\providecommand \doibase [0]{https://doi.org/}%
\providecommand \selectlanguage [0]{\@gobble}%
\providecommand \bibinfo  [0]{\@secondoftwo}%
\providecommand \bibfield  [0]{\@secondoftwo}%
\providecommand \translation [1]{[#1]}%
\providecommand \BibitemOpen [0]{}%
\providecommand \bibitemStop [0]{}%
\providecommand \bibitemNoStop [0]{.\EOS\space}%
\providecommand \EOS [0]{\spacefactor3000\relax}%
\providecommand \BibitemShut  [1]{\csname bibitem#1\endcsname}%
\let\auto@bib@innerbib\@empty
\bibitem [{\citenamefont {Ashcroft}\ and\ \citenamefont
  {Mermin}(1976)}]{neil1976solid}%
  \BibitemOpen
  \bibfield  {author} {\bibinfo {author} {\bibfnamefont {N.~W.}\ \bibnamefont
  {Ashcroft}}\ and\ \bibinfo {author} {\bibfnamefont {N.~D.}\ \bibnamefont
  {Mermin}},\ }\href@noop {} {\emph {\bibinfo {title} {Solid state physics}}}\
  (\bibinfo  {publisher} {Brooks/Cole, Cengage Learning},\ \bibinfo {year}
  {1976})\BibitemShut {NoStop}%
\bibitem [{\citenamefont {Zwanenburg}\ \emph {et~al.}(2013)\citenamefont
  {Zwanenburg}, \citenamefont {Dzurak}, \citenamefont {Morello}, \citenamefont
  {Simmons}, \citenamefont {Hollenberg}, \citenamefont {Klimeck}, \citenamefont
  {Rogge}, \citenamefont {Coppersmith},\ and\ \citenamefont
  {Eriksson}}]{ZwanenburgRMP-2013}%
  \BibitemOpen
  \bibfield  {author} {\bibinfo {author} {\bibfnamefont {F.~A.}\ \bibnamefont
  {Zwanenburg}}, \bibinfo {author} {\bibfnamefont {A.~S.}\ \bibnamefont
  {Dzurak}}, \bibinfo {author} {\bibfnamefont {A.}~\bibnamefont {Morello}},
  \bibinfo {author} {\bibfnamefont {M.~Y.}\ \bibnamefont {Simmons}}, \bibinfo
  {author} {\bibfnamefont {L.~C.~L.}\ \bibnamefont {Hollenberg}}, \bibinfo
  {author} {\bibfnamefont {G.}~\bibnamefont {Klimeck}}, \bibinfo {author}
  {\bibfnamefont {S.}~\bibnamefont {Rogge}}, \bibinfo {author} {\bibfnamefont
  {S.~N.}\ \bibnamefont {Coppersmith}},\ and\ \bibinfo {author} {\bibfnamefont
  {M.~A.}\ \bibnamefont {Eriksson}},\ }\bibfield  {title} {\bibinfo {title}
  {Silicon quantum electronics},\ }\href
  {https://doi.org/10.1103/RevModPhys.85.961} {\bibfield  {journal} {\bibinfo
  {journal} {Rev. Mod. Phys.}\ }\textbf {\bibinfo {volume} {85}},\ \bibinfo
  {pages} {961} (\bibinfo {year} {2013})}\BibitemShut {NoStop}%
\bibitem [{\citenamefont {{\v Z}uti{\'c}}\ \emph {et~al.}(2004)\citenamefont
  {{\v Z}uti{\'c}}, \citenamefont {Fabian},\ and\ \citenamefont
  {Das~Sarma}}]{zutic2004spintronics}%
  \BibitemOpen
  \bibfield  {author} {\bibinfo {author} {\bibfnamefont {I.}~\bibnamefont {{\v
  Z}uti{\'c}}}, \bibinfo {author} {\bibfnamefont {J.}~\bibnamefont {Fabian}},\
  and\ \bibinfo {author} {\bibfnamefont {S.}~\bibnamefont {Das~Sarma}},\
  }\bibfield  {title} {\bibinfo {title} {Spintronics: {{Fundamentals}} and
  applications},\ }\href {https://doi.org/10.1103/RevModPhys.76.323} {\bibfield
   {journal} {\bibinfo  {journal} {Reviews of Modern Physics}\ }\textbf
  {\bibinfo {volume} {76}},\ \bibinfo {pages} {323} (\bibinfo {year}
  {2004})}\BibitemShut {NoStop}%
\bibitem [{\citenamefont {Baltz}\ \emph {et~al.}(2018)\citenamefont {Baltz},
  \citenamefont {Manchon}, \citenamefont {Tsoi}, \citenamefont {Moriyama},
  \citenamefont {Ono},\ and\ \citenamefont
  {Tserkovnyak}}]{baltz2018antiferromagnetic}%
  \BibitemOpen
  \bibfield  {author} {\bibinfo {author} {\bibfnamefont {V.}~\bibnamefont
  {Baltz}}, \bibinfo {author} {\bibfnamefont {A.}~\bibnamefont {Manchon}},
  \bibinfo {author} {\bibfnamefont {M.}~\bibnamefont {Tsoi}}, \bibinfo {author}
  {\bibfnamefont {T.}~\bibnamefont {Moriyama}}, \bibinfo {author}
  {\bibfnamefont {T.}~\bibnamefont {Ono}},\ and\ \bibinfo {author}
  {\bibfnamefont {Y.}~\bibnamefont {Tserkovnyak}},\ }\bibfield  {title}
  {\bibinfo {title} {Antiferromagnetic spintronics},\ }\href
  {https://doi.org/10.1103/RevModPhys.90.015005} {\bibfield  {journal}
  {\bibinfo  {journal} {Reviews of Modern Physics}\ }\textbf {\bibinfo {volume}
  {90}},\ \bibinfo {pages} {015005} (\bibinfo {year} {2018})}\BibitemShut
  {NoStop}%
\bibitem [{\citenamefont {Schaibley}\ \emph {et~al.}(2016)\citenamefont
  {Schaibley}, \citenamefont {Yu}, \citenamefont {Clark}, \citenamefont
  {Rivera}, \citenamefont {Ross}, \citenamefont {Seyler}, \citenamefont {Yao},\
  and\ \citenamefont {Xu}}]{schaibley2016valleytronics}%
  \BibitemOpen
  \bibfield  {author} {\bibinfo {author} {\bibfnamefont {J.~R.}\ \bibnamefont
  {Schaibley}}, \bibinfo {author} {\bibfnamefont {H.}~\bibnamefont {Yu}},
  \bibinfo {author} {\bibfnamefont {G.}~\bibnamefont {Clark}}, \bibinfo
  {author} {\bibfnamefont {P.}~\bibnamefont {Rivera}}, \bibinfo {author}
  {\bibfnamefont {J.~S.}\ \bibnamefont {Ross}}, \bibinfo {author}
  {\bibfnamefont {K.~L.}\ \bibnamefont {Seyler}}, \bibinfo {author}
  {\bibfnamefont {W.}~\bibnamefont {Yao}},\ and\ \bibinfo {author}
  {\bibfnamefont {X.}~\bibnamefont {Xu}},\ }\bibfield  {title} {\bibinfo
  {title} {Valleytronics in {{2D}} materials},\ }\href
  {https://doi.org/10.1038/natrevmats.2016.55} {\bibfield  {journal} {\bibinfo
  {journal} {Nature Reviews Materials}\ }\textbf {\bibinfo {volume} {1}},\
  \bibinfo {pages} {1} (\bibinfo {year} {2016})}\BibitemShut {NoStop}%
\bibitem [{\citenamefont {Joachim}\ \emph {et~al.}(2000)\citenamefont
  {Joachim}, \citenamefont {Gimzewski},\ and\ \citenamefont
  {Aviram}}]{joachim2000electronics}%
  \BibitemOpen
  \bibfield  {author} {\bibinfo {author} {\bibfnamefont {C.}~\bibnamefont
  {Joachim}}, \bibinfo {author} {\bibfnamefont {J.~K.}\ \bibnamefont
  {Gimzewski}},\ and\ \bibinfo {author} {\bibfnamefont {A.}~\bibnamefont
  {Aviram}},\ }\bibfield  {title} {\bibinfo {title} {Electronics using
  hybrid-molecular and mono-molecular devices},\ }\href
  {https://doi.org/10.1038/35046000} {\bibfield  {journal} {\bibinfo  {journal}
  {Nature}\ }\textbf {\bibinfo {volume} {408}},\ \bibinfo {pages} {541}
  (\bibinfo {year} {2000})}\BibitemShut {NoStop}%
\bibitem [{\citenamefont {Wu}\ \emph {et~al.}(2007)\citenamefont {Wu},
  \citenamefont {Pisula},\ and\ \citenamefont {M{\"u}llen}}]{wu2007graphenes}%
  \BibitemOpen
  \bibfield  {author} {\bibinfo {author} {\bibfnamefont {J.}~\bibnamefont
  {Wu}}, \bibinfo {author} {\bibfnamefont {W.}~\bibnamefont {Pisula}},\ and\
  \bibinfo {author} {\bibfnamefont {K.}~\bibnamefont {M{\"u}llen}},\ }\bibfield
   {title} {\bibinfo {title} {Graphenes as {{Potential Material}} for
  {{Electronics}}},\ }\href {https://doi.org/10.1021/cr068010r} {\bibfield
  {journal} {\bibinfo  {journal} {Chemical Reviews}\ }\textbf {\bibinfo
  {volume} {107}},\ \bibinfo {pages} {718} (\bibinfo {year}
  {2007})}\BibitemShut {NoStop}%
\bibitem [{\citenamefont {Wang}\ \emph {et~al.}(2012)\citenamefont {Wang},
  \citenamefont {{Kalantar-Zadeh}}, \citenamefont {Kis}, \citenamefont
  {Coleman},\ and\ \citenamefont {Strano}}]{wang2012electronics}%
  \BibitemOpen
  \bibfield  {author} {\bibinfo {author} {\bibfnamefont {Q.~H.}\ \bibnamefont
  {Wang}}, \bibinfo {author} {\bibfnamefont {K.}~\bibnamefont
  {{Kalantar-Zadeh}}}, \bibinfo {author} {\bibfnamefont {A.}~\bibnamefont
  {Kis}}, \bibinfo {author} {\bibfnamefont {J.~N.}\ \bibnamefont {Coleman}},\
  and\ \bibinfo {author} {\bibfnamefont {M.~S.}\ \bibnamefont {Strano}},\
  }\bibfield  {title} {\bibinfo {title} {Electronics and optoelectronics of
  two-dimensional transition metal dichalcogenides},\ }\href
  {https://doi.org/10.1038/nnano.2012.193} {\bibfield  {journal} {\bibinfo
  {journal} {Nature Nanotechnology}\ }\textbf {\bibinfo {volume} {7}},\
  \bibinfo {pages} {699} (\bibinfo {year} {2012})}\BibitemShut {NoStop}%
\bibitem [{\citenamefont {Fiori}\ \emph {et~al.}(2014)\citenamefont {Fiori},
  \citenamefont {Bonaccorso}, \citenamefont {Iannaccone}, \citenamefont
  {Palacios}, \citenamefont {Neumaier}, \citenamefont {Seabaugh}, \citenamefont
  {Banerjee},\ and\ \citenamefont {Colombo}}]{fiori2014electronics}%
  \BibitemOpen
  \bibfield  {author} {\bibinfo {author} {\bibfnamefont {G.}~\bibnamefont
  {Fiori}}, \bibinfo {author} {\bibfnamefont {F.}~\bibnamefont {Bonaccorso}},
  \bibinfo {author} {\bibfnamefont {G.}~\bibnamefont {Iannaccone}}, \bibinfo
  {author} {\bibfnamefont {T.}~\bibnamefont {Palacios}}, \bibinfo {author}
  {\bibfnamefont {D.}~\bibnamefont {Neumaier}}, \bibinfo {author}
  {\bibfnamefont {A.}~\bibnamefont {Seabaugh}}, \bibinfo {author}
  {\bibfnamefont {S.~K.}\ \bibnamefont {Banerjee}},\ and\ \bibinfo {author}
  {\bibfnamefont {L.}~\bibnamefont {Colombo}},\ }\bibfield  {title} {\bibinfo
  {title} {Electronics based on two-dimensional materials},\ }\href
  {https://doi.org/10.1038/nnano.2014.207} {\bibfield  {journal} {\bibinfo
  {journal} {Nature Nanotechnology}\ }\textbf {\bibinfo {volume} {9}},\
  \bibinfo {pages} {768} (\bibinfo {year} {2014})}\BibitemShut {NoStop}%
\bibitem [{\citenamefont {Bader}\ and\ \citenamefont
  {Parkin}(2010)}]{bader2010spintronics}%
  \BibitemOpen
  \bibfield  {author} {\bibinfo {author} {\bibfnamefont {S.}~\bibnamefont
  {Bader}}\ and\ \bibinfo {author} {\bibfnamefont {S.}~\bibnamefont {Parkin}},\
  }\bibfield  {title} {\bibinfo {title} {Spintronics},\ }\href
  {https://doi.org/10.1146/annurev-conmatphys-070909-104123} {\bibfield
  {journal} {\bibinfo  {journal} {Annual Review of Condensed Matter Physics}\
  }\textbf {\bibinfo {volume} {1}},\ \bibinfo {pages} {71} (\bibinfo {year}
  {2010})}\BibitemShut {NoStop}%
\bibitem [{\citenamefont {Wolf}\ \emph {et~al.}(2001)\citenamefont {Wolf},
  \citenamefont {Awschalom}, \citenamefont {Buhrman}, \citenamefont {Daughton},
  \citenamefont {{von Moln{\'a}r}}, \citenamefont {Roukes}, \citenamefont
  {Chtchelkanova},\ and\ \citenamefont {Treger}}]{wolf2001spintronics}%
  \BibitemOpen
  \bibfield  {author} {\bibinfo {author} {\bibfnamefont {S.~A.}\ \bibnamefont
  {Wolf}}, \bibinfo {author} {\bibfnamefont {D.~D.}\ \bibnamefont {Awschalom}},
  \bibinfo {author} {\bibfnamefont {R.~A.}\ \bibnamefont {Buhrman}}, \bibinfo
  {author} {\bibfnamefont {J.~M.}\ \bibnamefont {Daughton}}, \bibinfo {author}
  {\bibfnamefont {S.}~\bibnamefont {{von Moln{\'a}r}}}, \bibinfo {author}
  {\bibfnamefont {M.~L.}\ \bibnamefont {Roukes}}, \bibinfo {author}
  {\bibfnamefont {A.~Y.}\ \bibnamefont {Chtchelkanova}},\ and\ \bibinfo
  {author} {\bibfnamefont {D.~M.}\ \bibnamefont {Treger}},\ }\bibfield  {title}
  {\bibinfo {title} {Spintronics: {{A Spin-Based Electronics Vision}} for the
  {{Future}}},\ }\href {https://doi.org/10.1126/science.1065389} {\bibfield
  {journal} {\bibinfo  {journal} {Science}\ }\textbf {\bibinfo {volume}
  {294}},\ \bibinfo {pages} {1488} (\bibinfo {year} {2001})}\BibitemShut
  {NoStop}%
\bibitem [{\citenamefont {Chappert}\ \emph {et~al.}(2007)\citenamefont
  {Chappert}, \citenamefont {Fert},\ and\ \citenamefont
  {Van~Dau}}]{chappert2007emergence}%
  \BibitemOpen
  \bibfield  {author} {\bibinfo {author} {\bibfnamefont {C.}~\bibnamefont
  {Chappert}}, \bibinfo {author} {\bibfnamefont {A.}~\bibnamefont {Fert}},\
  and\ \bibinfo {author} {\bibfnamefont {F.~N.}\ \bibnamefont {Van~Dau}},\
  }\bibfield  {title} {\bibinfo {title} {The emergence of spin electronics in
  data storage},\ }\href {https://doi.org/10.1038/nmat2024} {\bibfield
  {journal} {\bibinfo  {journal} {Nature Materials}\ }\textbf {\bibinfo
  {volume} {6}},\ \bibinfo {pages} {813} (\bibinfo {year} {2007})}\BibitemShut
  {NoStop}%
\bibitem [{\citenamefont {Awschalom}\ and\ \citenamefont
  {Flatt{\'e}}(2007)}]{awschalom2007challenges}%
  \BibitemOpen
  \bibfield  {author} {\bibinfo {author} {\bibfnamefont {D.~D.}\ \bibnamefont
  {Awschalom}}\ and\ \bibinfo {author} {\bibfnamefont {M.~E.}\ \bibnamefont
  {Flatt{\'e}}},\ }\bibfield  {title} {\bibinfo {title} {Challenges for
  semiconductor spintronics},\ }\href {https://doi.org/10.1038/nphys551}
  {\bibfield  {journal} {\bibinfo  {journal} {Nature Physics}\ }\textbf
  {\bibinfo {volume} {3}},\ \bibinfo {pages} {153} (\bibinfo {year}
  {2007})}\BibitemShut {NoStop}%
\bibitem [{\citenamefont {Rycerz}\ \emph {et~al.}(2007)\citenamefont {Rycerz},
  \citenamefont {Tworzyd{\l}o},\ and\ \citenamefont
  {Beenakker}}]{rycerz2007valley}%
  \BibitemOpen
  \bibfield  {author} {\bibinfo {author} {\bibfnamefont {A.}~\bibnamefont
  {Rycerz}}, \bibinfo {author} {\bibfnamefont {J.}~\bibnamefont
  {Tworzyd{\l}o}},\ and\ \bibinfo {author} {\bibfnamefont {C.~W.~J.}\
  \bibnamefont {Beenakker}},\ }\bibfield  {title} {\bibinfo {title} {Valley
  filter and valley valve in graphene},\ }\href
  {https://doi.org/10.1038/nphys547} {\bibfield  {journal} {\bibinfo  {journal}
  {Nature Physics}\ }\textbf {\bibinfo {volume} {3}},\ \bibinfo {pages} {172}
  (\bibinfo {year} {2007})}\BibitemShut {NoStop}%
\bibitem [{\citenamefont {Xu}\ \emph {et~al.}(2014)\citenamefont {Xu},
  \citenamefont {Yao}, \citenamefont {Xiao},\ and\ \citenamefont
  {Heinz}}]{xu2014spin}%
  \BibitemOpen
  \bibfield  {author} {\bibinfo {author} {\bibfnamefont {X.}~\bibnamefont
  {Xu}}, \bibinfo {author} {\bibfnamefont {W.}~\bibnamefont {Yao}}, \bibinfo
  {author} {\bibfnamefont {D.}~\bibnamefont {Xiao}},\ and\ \bibinfo {author}
  {\bibfnamefont {T.~F.}\ \bibnamefont {Heinz}},\ }\bibfield  {title} {\bibinfo
  {title} {Spin and pseudospins in layered transition metal dichalcogenides},\
  }\href {https://doi.org/10.1038/nphys2942} {\bibfield  {journal} {\bibinfo
  {journal} {Nature Physics}\ }\textbf {\bibinfo {volume} {10}},\ \bibinfo
  {pages} {343} (\bibinfo {year} {2014})}\BibitemShut {NoStop}%
\bibitem [{\citenamefont {Sui}\ \emph {et~al.}(2015)\citenamefont {Sui},
  \citenamefont {Chen}, \citenamefont {Ma}, \citenamefont {Shan}, \citenamefont
  {Tian}, \citenamefont {Watanabe}, \citenamefont {Taniguchi}, \citenamefont
  {Jin}, \citenamefont {Yao}, \citenamefont {Xiao},\ and\ \citenamefont
  {Zhang}}]{sui2015gatetunable}%
  \BibitemOpen
  \bibfield  {author} {\bibinfo {author} {\bibfnamefont {M.}~\bibnamefont
  {Sui}}, \bibinfo {author} {\bibfnamefont {G.}~\bibnamefont {Chen}}, \bibinfo
  {author} {\bibfnamefont {L.}~\bibnamefont {Ma}}, \bibinfo {author}
  {\bibfnamefont {W.-Y.}\ \bibnamefont {Shan}}, \bibinfo {author}
  {\bibfnamefont {D.}~\bibnamefont {Tian}}, \bibinfo {author} {\bibfnamefont
  {K.}~\bibnamefont {Watanabe}}, \bibinfo {author} {\bibfnamefont
  {T.}~\bibnamefont {Taniguchi}}, \bibinfo {author} {\bibfnamefont
  {X.}~\bibnamefont {Jin}}, \bibinfo {author} {\bibfnamefont {W.}~\bibnamefont
  {Yao}}, \bibinfo {author} {\bibfnamefont {D.}~\bibnamefont {Xiao}},\ and\
  \bibinfo {author} {\bibfnamefont {Y.}~\bibnamefont {Zhang}},\ }\bibfield
  {title} {\bibinfo {title} {Gate-tunable topological valley transport in
  bilayer graphene},\ }\href {https://doi.org/10.1038/nphys3485} {\bibfield
  {journal} {\bibinfo  {journal} {Nature Physics}\ }\textbf {\bibinfo {volume}
  {11}},\ \bibinfo {pages} {1027} (\bibinfo {year} {2015})}\BibitemShut
  {NoStop}%
\bibitem [{\citenamefont {Tong}\ \emph {et~al.}(2016)\citenamefont {Tong},
  \citenamefont {Gong}, \citenamefont {Wan},\ and\ \citenamefont
  {Duan}}]{tong2016concepts}%
  \BibitemOpen
  \bibfield  {author} {\bibinfo {author} {\bibfnamefont {W.-Y.}\ \bibnamefont
  {Tong}}, \bibinfo {author} {\bibfnamefont {S.-J.}\ \bibnamefont {Gong}},
  \bibinfo {author} {\bibfnamefont {X.}~\bibnamefont {Wan}},\ and\ \bibinfo
  {author} {\bibfnamefont {C.-G.}\ \bibnamefont {Duan}},\ }\bibfield  {title}
  {\bibinfo {title} {Concepts of ferrovalley material and anomalous valley
  {{Hall}} effect},\ }\href {https://doi.org/10.1038/ncomms13612} {\bibfield
  {journal} {\bibinfo  {journal} {Nature Communications}\ }\textbf {\bibinfo
  {volume} {7}},\ \bibinfo {pages} {13612} (\bibinfo {year}
  {2016})}\BibitemShut {NoStop}%
\bibitem [{\citenamefont {Yu}\ \emph {et~al.}(2020)\citenamefont {Yu},
  \citenamefont {Guan}, \citenamefont {Sheng}, \citenamefont {Gao},\ and\
  \citenamefont {Yang}}]{yu2020valleylayer}%
  \BibitemOpen
  \bibfield  {author} {\bibinfo {author} {\bibfnamefont {Z.-M.}\ \bibnamefont
  {Yu}}, \bibinfo {author} {\bibfnamefont {S.}~\bibnamefont {Guan}}, \bibinfo
  {author} {\bibfnamefont {X.-L.}\ \bibnamefont {Sheng}}, \bibinfo {author}
  {\bibfnamefont {W.}~\bibnamefont {Gao}},\ and\ \bibinfo {author}
  {\bibfnamefont {S.~A.}\ \bibnamefont {Yang}},\ }\bibfield  {title} {\bibinfo
  {title} {Valley-{{Layer Coupling}}: {{A New Design Principle}} for
  {{Valleytronics}}},\ }\href {https://doi.org/10.1103/PhysRevLett.124.037701}
  {\bibfield  {journal} {\bibinfo  {journal} {Physical Review Letters}\
  }\textbf {\bibinfo {volume} {124}},\ \bibinfo {pages} {037701} (\bibinfo
  {year} {2020})}\BibitemShut {NoStop}%
\bibitem [{\citenamefont {Xiao}\ \emph {et~al.}(2012)\citenamefont {Xiao},
  \citenamefont {Liu}, \citenamefont {Feng}, \citenamefont {Xu},\ and\
  \citenamefont {Yao}}]{xiao2012coupleda}%
  \BibitemOpen
  \bibfield  {author} {\bibinfo {author} {\bibfnamefont {D.}~\bibnamefont
  {Xiao}}, \bibinfo {author} {\bibfnamefont {G.-B.}\ \bibnamefont {Liu}},
  \bibinfo {author} {\bibfnamefont {W.}~\bibnamefont {Feng}}, \bibinfo {author}
  {\bibfnamefont {X.}~\bibnamefont {Xu}},\ and\ \bibinfo {author}
  {\bibfnamefont {W.}~\bibnamefont {Yao}},\ }\bibfield  {title} {\bibinfo
  {title} {Coupled {{Spin}} and {{Valley Physics}} in {{Monolayers}} of
  {{MoS}}${}_2$ and {{Other Group-VI Dichalcogenides}}},\ }\href
  {https://doi.org/10.1103/PhysRevLett.108.196802} {\bibfield  {journal}
  {\bibinfo  {journal} {Physical Review Letters}\ }\textbf {\bibinfo {volume}
  {108}},\ \bibinfo {pages} {196802} (\bibinfo {year} {2012})}\BibitemShut
  {NoStop}%
\bibitem [{\citenamefont {Zhu}\ \emph {et~al.}(2012)\citenamefont {Zhu},
  \citenamefont {Collaudin}, \citenamefont {Fauqu{\'e}}, \citenamefont {Kang},\
  and\ \citenamefont {Behnia}}]{zhu2012fieldinduced}%
  \BibitemOpen
  \bibfield  {author} {\bibinfo {author} {\bibfnamefont {Z.}~\bibnamefont
  {Zhu}}, \bibinfo {author} {\bibfnamefont {A.}~\bibnamefont {Collaudin}},
  \bibinfo {author} {\bibfnamefont {B.}~\bibnamefont {Fauqu{\'e}}}, \bibinfo
  {author} {\bibfnamefont {W.}~\bibnamefont {Kang}},\ and\ \bibinfo {author}
  {\bibfnamefont {K.}~\bibnamefont {Behnia}},\ }\bibfield  {title} {\bibinfo
  {title} {Field-induced polarization of {{Dirac}} valleys in bismuth},\ }\href
  {https://doi.org/10.1038/nphys2111} {\bibfield  {journal} {\bibinfo
  {journal} {Nature Physics}\ }\textbf {\bibinfo {volume} {8}},\ \bibinfo
  {pages} {89} (\bibinfo {year} {2012})}\BibitemShut {NoStop}%
\bibitem [{\citenamefont {Jiang}\ \emph {et~al.}(2013)\citenamefont {Jiang},
  \citenamefont {Low}, \citenamefont {Chang}, \citenamefont {Katsnelson},\ and\
  \citenamefont {Guinea}}]{jiang2013generation}%
  \BibitemOpen
  \bibfield  {author} {\bibinfo {author} {\bibfnamefont {Y.}~\bibnamefont
  {Jiang}}, \bibinfo {author} {\bibfnamefont {T.}~\bibnamefont {Low}}, \bibinfo
  {author} {\bibfnamefont {K.}~\bibnamefont {Chang}}, \bibinfo {author}
  {\bibfnamefont {M.~I.}\ \bibnamefont {Katsnelson}},\ and\ \bibinfo {author}
  {\bibfnamefont {F.}~\bibnamefont {Guinea}},\ }\bibfield  {title} {\bibinfo
  {title} {Generation of {{Pure Bulk Valley Current}} in {{Graphene}}},\ }\href
  {https://doi.org/10.1103/PhysRevLett.110.046601} {\bibfield  {journal}
  {\bibinfo  {journal} {Physical Review Letters}\ }\textbf {\bibinfo {volume}
  {110}},\ \bibinfo {pages} {046601} (\bibinfo {year} {2013})}\BibitemShut
  {NoStop}%
\bibitem [{\citenamefont {Gruji{\'c}}\ \emph {et~al.}(2014)\citenamefont
  {Gruji{\'c}}, \citenamefont {Tadi{\'c}},\ and\ \citenamefont
  {Peeters}}]{grujic2014spinvalley}%
  \BibitemOpen
  \bibfield  {author} {\bibinfo {author} {\bibfnamefont {M.~M.}\ \bibnamefont
  {Gruji{\'c}}}, \bibinfo {author} {\bibfnamefont {M.~{\v Z}.}\ \bibnamefont
  {Tadi{\'c}}},\ and\ \bibinfo {author} {\bibfnamefont {F.~M.}\ \bibnamefont
  {Peeters}},\ }\bibfield  {title} {\bibinfo {title} {Spin-{{Valley Filtering}}
  in {{Strained Graphene Structures}} with {{Artificially Induced Carrier
  Mass}} and {{Spin-Orbit Coupling}}},\ }\href
  {https://doi.org/10.1103/PhysRevLett.113.046601} {\bibfield  {journal}
  {\bibinfo  {journal} {Physical Review Letters}\ }\textbf {\bibinfo {volume}
  {113}},\ \bibinfo {pages} {046601} (\bibinfo {year} {2014})}\BibitemShut
  {NoStop}%
\bibitem [{\citenamefont {Qiao}\ \emph {et~al.}(2014)\citenamefont {Qiao},
  \citenamefont {Jung}, \citenamefont {Lin}, \citenamefont {Ren}, \citenamefont
  {MacDonald},\ and\ \citenamefont {Niu}}]{qiao2014current}%
  \BibitemOpen
  \bibfield  {author} {\bibinfo {author} {\bibfnamefont {Z.}~\bibnamefont
  {Qiao}}, \bibinfo {author} {\bibfnamefont {J.}~\bibnamefont {Jung}}, \bibinfo
  {author} {\bibfnamefont {C.}~\bibnamefont {Lin}}, \bibinfo {author}
  {\bibfnamefont {Y.}~\bibnamefont {Ren}}, \bibinfo {author} {\bibfnamefont
  {A.~H.}\ \bibnamefont {MacDonald}},\ and\ \bibinfo {author} {\bibfnamefont
  {Q.}~\bibnamefont {Niu}},\ }\bibfield  {title} {\bibinfo {title} {Current
  {{Partition}} at {{Topological Channel Intersections}}},\ }\href
  {https://doi.org/10.1103/PhysRevLett.112.206601} {\bibfield  {journal}
  {\bibinfo  {journal} {Physical Review Letters}\ }\textbf {\bibinfo {volume}
  {112}},\ \bibinfo {pages} {206601} (\bibinfo {year} {2014})}\BibitemShut
  {NoStop}%
\bibitem [{\citenamefont {Pan}\ \emph {et~al.}(2015{\natexlab{a}})\citenamefont
  {Pan}, \citenamefont {Li}, \citenamefont {Jiang}, \citenamefont {Yao},\ and\
  \citenamefont {Yang}}]{pan2015valleypolarized}%
  \BibitemOpen
  \bibfield  {author} {\bibinfo {author} {\bibfnamefont {H.}~\bibnamefont
  {Pan}}, \bibinfo {author} {\bibfnamefont {X.}~\bibnamefont {Li}}, \bibinfo
  {author} {\bibfnamefont {H.}~\bibnamefont {Jiang}}, \bibinfo {author}
  {\bibfnamefont {Y.}~\bibnamefont {Yao}},\ and\ \bibinfo {author}
  {\bibfnamefont {S.~A.}\ \bibnamefont {Yang}},\ }\bibfield  {title} {\bibinfo
  {title} {Valley-polarized quantum anomalous {{Hall}} phase and
  disorder-induced valley-filtered chiral edge channels},\ }\href
  {https://doi.org/10.1103/PhysRevB.91.045404} {\bibfield  {journal} {\bibinfo
  {journal} {Physical Review B}\ }\textbf {\bibinfo {volume} {91}},\ \bibinfo
  {pages} {045404} (\bibinfo {year} {2015}{\natexlab{a}})}\BibitemShut
  {NoStop}%
\bibitem [{\citenamefont {Pan}\ \emph {et~al.}(2015{\natexlab{b}})\citenamefont
  {Pan}, \citenamefont {Li}, \citenamefont {Zhang},\ and\ \citenamefont
  {Yang}}]{pan2015perfect}%
  \BibitemOpen
  \bibfield  {author} {\bibinfo {author} {\bibfnamefont {H.}~\bibnamefont
  {Pan}}, \bibinfo {author} {\bibfnamefont {X.}~\bibnamefont {Li}}, \bibinfo
  {author} {\bibfnamefont {F.}~\bibnamefont {Zhang}},\ and\ \bibinfo {author}
  {\bibfnamefont {S.~A.}\ \bibnamefont {Yang}},\ }\bibfield  {title} {\bibinfo
  {title} {Perfect valley filter in a topological domain wall},\ }\href
  {https://doi.org/10.1103/PhysRevB.92.041404} {\bibfield  {journal} {\bibinfo
  {journal} {Physical Review B}\ }\textbf {\bibinfo {volume} {92}},\ \bibinfo
  {pages} {041404} (\bibinfo {year} {2015}{\natexlab{b}})}\BibitemShut
  {NoStop}%
\bibitem [{\citenamefont {Settnes}\ \emph {et~al.}(2016)\citenamefont
  {Settnes}, \citenamefont {Power}, \citenamefont {Brandbyge},\ and\
  \citenamefont {Jauho}}]{settnes2016graphene}%
  \BibitemOpen
  \bibfield  {author} {\bibinfo {author} {\bibfnamefont {M.}~\bibnamefont
  {Settnes}}, \bibinfo {author} {\bibfnamefont {S.~R.}\ \bibnamefont {Power}},
  \bibinfo {author} {\bibfnamefont {M.}~\bibnamefont {Brandbyge}},\ and\
  \bibinfo {author} {\bibfnamefont {A.-P.}\ \bibnamefont {Jauho}},\ }\bibfield
  {title} {\bibinfo {title} {Graphene {{Nanobubbles}} as {{Valley Filters}} and
  {{Beam Splitters}}},\ }\href {https://doi.org/10.1103/PhysRevLett.117.276801}
  {\bibfield  {journal} {\bibinfo  {journal} {Physical Review Letters}\
  }\textbf {\bibinfo {volume} {117}},\ \bibinfo {pages} {276801} (\bibinfo
  {year} {2016})}\BibitemShut {NoStop}%
\bibitem [{\citenamefont {Cheng}\ \emph {et~al.}(2018)\citenamefont {Cheng},
  \citenamefont {Liu}, \citenamefont {Jiang}, \citenamefont {Sun},\ and\
  \citenamefont {Xie}}]{cheng2018manipulation}%
  \BibitemOpen
  \bibfield  {author} {\bibinfo {author} {\bibfnamefont {S.-g.}\ \bibnamefont
  {Cheng}}, \bibinfo {author} {\bibfnamefont {H.}~\bibnamefont {Liu}}, \bibinfo
  {author} {\bibfnamefont {H.}~\bibnamefont {Jiang}}, \bibinfo {author}
  {\bibfnamefont {Q.-F.}\ \bibnamefont {Sun}},\ and\ \bibinfo {author}
  {\bibfnamefont {X.~C.}\ \bibnamefont {Xie}},\ }\bibfield  {title} {\bibinfo
  {title} {Manipulation and {{Characterization}} of the {{Valley-Polarized
  Topological Kink States}} in {{Graphene-Based Interferometers}}},\ }\href
  {https://doi.org/10.1103/PhysRevLett.121.156801} {\bibfield  {journal}
  {\bibinfo  {journal} {Physical Review Letters}\ }\textbf {\bibinfo {volume}
  {121}},\ \bibinfo {pages} {156801} (\bibinfo {year} {2018})}\BibitemShut
  {NoStop}%
\bibitem [{\citenamefont {Li}\ \emph {et~al.}(2018)\citenamefont {Li},
  \citenamefont {Zhang}, \citenamefont {Yin}, \citenamefont {Zhang},
  \citenamefont {Watanabe}, \citenamefont {Taniguchi}, \citenamefont {Liu},\
  and\ \citenamefont {Zhu}}]{li2018valley}%
  \BibitemOpen
  \bibfield  {author} {\bibinfo {author} {\bibfnamefont {J.}~\bibnamefont
  {Li}}, \bibinfo {author} {\bibfnamefont {R.-X.}\ \bibnamefont {Zhang}},
  \bibinfo {author} {\bibfnamefont {Z.}~\bibnamefont {Yin}}, \bibinfo {author}
  {\bibfnamefont {J.}~\bibnamefont {Zhang}}, \bibinfo {author} {\bibfnamefont
  {K.}~\bibnamefont {Watanabe}}, \bibinfo {author} {\bibfnamefont
  {T.}~\bibnamefont {Taniguchi}}, \bibinfo {author} {\bibfnamefont
  {C.}~\bibnamefont {Liu}},\ and\ \bibinfo {author} {\bibfnamefont
  {J.}~\bibnamefont {Zhu}},\ }\bibfield  {title} {\bibinfo {title} {A valley
  valve and electron beam splitter},\ }\href
  {https://doi.org/10.1126/science.aao5989} {\bibfield  {journal} {\bibinfo
  {journal} {Science}\ }\textbf {\bibinfo {volume} {362}},\ \bibinfo {pages}
  {1149} (\bibinfo {year} {2018})}\BibitemShut {NoStop}%
\bibitem [{\citenamefont {Benalcazar}\ \emph {et~al.}(2017)\citenamefont
  {Benalcazar}, \citenamefont {Bernevig},\ and\ \citenamefont
  {Hughes}}]{benalcazar2017quantized}%
  \BibitemOpen
  \bibfield  {author} {\bibinfo {author} {\bibfnamefont {W.~A.}\ \bibnamefont
  {Benalcazar}}, \bibinfo {author} {\bibfnamefont {B.~A.}\ \bibnamefont
  {Bernevig}},\ and\ \bibinfo {author} {\bibfnamefont {T.~L.}\ \bibnamefont
  {Hughes}},\ }\bibfield  {title} {\bibinfo {title} {Quantized electric
  multipole insulators},\ }\href {https://doi.org/10.1126/science.aah6442}
  {\bibfield  {journal} {\bibinfo  {journal} {Science}\ }\textbf {\bibinfo
  {volume} {357}},\ \bibinfo {pages} {61} (\bibinfo {year} {2017})}\BibitemShut
  {NoStop}%
\bibitem [{\citenamefont {Langbehn}\ \emph {et~al.}(2017)\citenamefont
  {Langbehn}, \citenamefont {Peng}, \citenamefont {Trifunovic}, \citenamefont
  {{von Oppen}},\ and\ \citenamefont
  {Brouwer}}]{langbehn2017reflectionsymmetrica}%
  \BibitemOpen
  \bibfield  {author} {\bibinfo {author} {\bibfnamefont {J.}~\bibnamefont
  {Langbehn}}, \bibinfo {author} {\bibfnamefont {Y.}~\bibnamefont {Peng}},
  \bibinfo {author} {\bibfnamefont {L.}~\bibnamefont {Trifunovic}}, \bibinfo
  {author} {\bibfnamefont {F.}~\bibnamefont {{von Oppen}}},\ and\ \bibinfo
  {author} {\bibfnamefont {P.~W.}\ \bibnamefont {Brouwer}},\ }\bibfield
  {title} {\bibinfo {title} {Reflection-{{Symmetric Second-Order Topological
  Insulators}} and {{Superconductors}}},\ }\href
  {https://doi.org/10.1103/PhysRevLett.119.246401} {\bibfield  {journal}
  {\bibinfo  {journal} {Physical Review Letters}\ }\textbf {\bibinfo {volume}
  {119}},\ \bibinfo {pages} {246401} (\bibinfo {year} {2017})}\BibitemShut
  {NoStop}%
\bibitem [{\citenamefont {Song}\ \emph {et~al.}(2017)\citenamefont {Song},
  \citenamefont {Fang},\ and\ \citenamefont {Fang}}]{song2017ensuremath2}%
  \BibitemOpen
  \bibfield  {author} {\bibinfo {author} {\bibfnamefont {Z.}~\bibnamefont
  {Song}}, \bibinfo {author} {\bibfnamefont {Z.}~\bibnamefont {Fang}},\ and\
  \bibinfo {author} {\bibfnamefont {C.}~\bibnamefont {Fang}},\ }\bibfield
  {title} {\bibinfo {title} {(d-2)-{{Dimensional Edge States}} of {{Rotation
  Symmetry Protected Topological States}}},\ }\href
  {https://doi.org/10.1103/PhysRevLett.119.246402} {\bibfield  {journal}
  {\bibinfo  {journal} {Physical Review Letters}\ }\textbf {\bibinfo {volume}
  {119}},\ \bibinfo {pages} {246402} (\bibinfo {year} {2017})}\BibitemShut
  {NoStop}%
\bibitem [{\citenamefont {Wang}\ \emph {et~al.}(2018)\citenamefont {Wang},
  \citenamefont {Liu}, \citenamefont {Lu},\ and\ \citenamefont
  {Zhang}}]{wang2018hightemperature}%
  \BibitemOpen
  \bibfield  {author} {\bibinfo {author} {\bibfnamefont {Q.}~\bibnamefont
  {Wang}}, \bibinfo {author} {\bibfnamefont {C.-C.}\ \bibnamefont {Liu}},
  \bibinfo {author} {\bibfnamefont {Y.-M.}\ \bibnamefont {Lu}},\ and\ \bibinfo
  {author} {\bibfnamefont {F.}~\bibnamefont {Zhang}},\ }\bibfield  {title}
  {\bibinfo {title} {High-{{Temperature Majorana Corner States}}},\ }\href
  {https://doi.org/10.1103/PhysRevLett.121.186801} {\bibfield  {journal}
  {\bibinfo  {journal} {Physical Review Letters}\ }\textbf {\bibinfo {volume}
  {121}},\ \bibinfo {pages} {186801} (\bibinfo {year} {2018})}\BibitemShut
  {NoStop}%
\bibitem [{\citenamefont {Schindler}\ \emph {et~al.}(2018)\citenamefont
  {Schindler}, \citenamefont {Cook}, \citenamefont {Vergniory}, \citenamefont
  {Wang}, \citenamefont {Parkin}, \citenamefont {Bernevig},\ and\ \citenamefont
  {Neupert}}]{schindler2018higherordera}%
  \BibitemOpen
  \bibfield  {author} {\bibinfo {author} {\bibfnamefont {F.}~\bibnamefont
  {Schindler}}, \bibinfo {author} {\bibfnamefont {A.~M.}\ \bibnamefont {Cook}},
  \bibinfo {author} {\bibfnamefont {M.~G.}\ \bibnamefont {Vergniory}}, \bibinfo
  {author} {\bibfnamefont {Z.}~\bibnamefont {Wang}}, \bibinfo {author}
  {\bibfnamefont {S.~S.~P.}\ \bibnamefont {Parkin}}, \bibinfo {author}
  {\bibfnamefont {B.~A.}\ \bibnamefont {Bernevig}},\ and\ \bibinfo {author}
  {\bibfnamefont {T.}~\bibnamefont {Neupert}},\ }\bibfield  {title} {\bibinfo
  {title} {Higher-order topological insulators},\ }\href
  {https://doi.org/10.1126/sciadv.aat0346} {\bibfield  {journal} {\bibinfo
  {journal} {Science Advances}\ }\textbf {\bibinfo {volume} {4}},\ \bibinfo
  {pages} {eaat0346} (\bibinfo {year} {2018})}\BibitemShut {NoStop}%
\bibitem [{\citenamefont {Ezawa}(2018{\natexlab{a}})}]{ezawa2018magnetic}%
  \BibitemOpen
  \bibfield  {author} {\bibinfo {author} {\bibfnamefont {M.}~\bibnamefont
  {Ezawa}},\ }\bibfield  {title} {\bibinfo {title} {Magnetic second-order
  topological insulators and semimetals},\ }\href
  {https://doi.org/10.1103/PhysRevB.97.155305} {\bibfield  {journal} {\bibinfo
  {journal} {Physical Review B}\ }\textbf {\bibinfo {volume} {97}},\ \bibinfo
  {pages} {155305} (\bibinfo {year} {2018}{\natexlab{a}})}\BibitemShut
  {NoStop}%
\bibitem [{\citenamefont {Ezawa}(2018{\natexlab{b}})}]{ezawa2018topological}%
  \BibitemOpen
  \bibfield  {author} {\bibinfo {author} {\bibfnamefont {M.}~\bibnamefont
  {Ezawa}},\ }\bibfield  {title} {\bibinfo {title} {Topological {{Switch}}
  between {{Second-Order Topological Insulators}} and {{Topological Crystalline
  Insulators}}},\ }\href {https://doi.org/10.1103/PhysRevLett.121.116801}
  {\bibfield  {journal} {\bibinfo  {journal} {Physical Review Letters}\
  }\textbf {\bibinfo {volume} {121}},\ \bibinfo {pages} {116801} (\bibinfo
  {year} {2018}{\natexlab{b}})}\BibitemShut {NoStop}%
\bibitem [{\citenamefont {Wu}\ \emph {et~al.}(2022)\citenamefont {Wu},
  \citenamefont {Guo}, \citenamefont {Zhang},\ and\ \citenamefont
  {Jiang}}]{wu2022quantized}%
  \BibitemOpen
  \bibfield  {author} {\bibinfo {author} {\bibfnamefont {B.-L.}\ \bibnamefont
  {Wu}}, \bibinfo {author} {\bibfnamefont {A.-M.}\ \bibnamefont {Guo}},
  \bibinfo {author} {\bibfnamefont {Z.-Q.}\ \bibnamefont {Zhang}},\ and\
  \bibinfo {author} {\bibfnamefont {H.}~\bibnamefont {Jiang}},\ }\bibfield
  {title} {\bibinfo {title} {Quantized charge-pumping in higher-order
  topological insulators},\ }\href
  {https://doi.org/10.1103/PhysRevB.106.165401} {\bibfield  {journal} {\bibinfo
   {journal} {Physical Review B}\ }\textbf {\bibinfo {volume} {106}},\ \bibinfo
  {pages} {165401} (\bibinfo {year} {2022})}\BibitemShut {NoStop}%
\bibitem [{\citenamefont {Sheng}\ \emph {et~al.}(2019)\citenamefont {Sheng},
  \citenamefont {Chen}, \citenamefont {Liu}, \citenamefont {Chen},
  \citenamefont {Yu}, \citenamefont {Zhao},\ and\ \citenamefont
  {Yang}}]{sheng2019twodimensional}%
  \BibitemOpen
  \bibfield  {author} {\bibinfo {author} {\bibfnamefont {X.-L.}\ \bibnamefont
  {Sheng}}, \bibinfo {author} {\bibfnamefont {C.}~\bibnamefont {Chen}},
  \bibinfo {author} {\bibfnamefont {H.}~\bibnamefont {Liu}}, \bibinfo {author}
  {\bibfnamefont {Z.}~\bibnamefont {Chen}}, \bibinfo {author} {\bibfnamefont
  {Z.-M.}\ \bibnamefont {Yu}}, \bibinfo {author} {\bibfnamefont {Y.~X.}\
  \bibnamefont {Zhao}},\ and\ \bibinfo {author} {\bibfnamefont {S.~A.}\
  \bibnamefont {Yang}},\ }\bibfield  {title} {\bibinfo {title}
  {Two-{{Dimensional Second-Order Topological Insulator}} in {{Graphdiyne}}},\
  }\href {https://doi.org/10.1103/PhysRevLett.123.256402} {\bibfield  {journal}
  {\bibinfo  {journal} {Physical Review Letters}\ }\textbf {\bibinfo {volume}
  {123}},\ \bibinfo {pages} {256402} (\bibinfo {year} {2019})}\BibitemShut
  {NoStop}%
\bibitem [{\citenamefont {Lee}\ \emph {et~al.}(2020)\citenamefont {Lee},
  \citenamefont {Kim}, \citenamefont {Ahn},\ and\ \citenamefont
  {Yang}}]{RN2830}%
  \BibitemOpen
  \bibfield  {author} {\bibinfo {author} {\bibfnamefont {E.}~\bibnamefont
  {Lee}}, \bibinfo {author} {\bibfnamefont {R.}~\bibnamefont {Kim}}, \bibinfo
  {author} {\bibfnamefont {J.}~\bibnamefont {Ahn}},\ and\ \bibinfo {author}
  {\bibfnamefont {B.-J.}\ \bibnamefont {Yang}},\ }\bibfield  {title} {\bibinfo
  {title} {Two-dimensional higher-order topology in monolayer graphdiyne},\
  }\bibfield  {journal} {\bibinfo  {journal} {npj Quantum Materials}\ }\textbf
  {\bibinfo {volume} {5}},\ \href {https://doi.org/10.1038/s41535-019-0206-8}
  {10.1038/s41535-019-0206-8} (\bibinfo {year} {2020})\BibitemShut {NoStop}%
\bibitem [{\citenamefont {Guo}\ \emph {et~al.}(2022)\citenamefont {Guo},
  \citenamefont {Deng}, \citenamefont {Xie},\ and\ \citenamefont
  {Wang}}]{guo2022quadrupole}%
  \BibitemOpen
  \bibfield  {author} {\bibinfo {author} {\bibfnamefont {Z.}~\bibnamefont
  {Guo}}, \bibinfo {author} {\bibfnamefont {J.}~\bibnamefont {Deng}}, \bibinfo
  {author} {\bibfnamefont {Y.}~\bibnamefont {Xie}},\ and\ \bibinfo {author}
  {\bibfnamefont {Z.}~\bibnamefont {Wang}},\ }\bibfield  {title} {\bibinfo
  {title} {Quadrupole topological insulators in
  {{Ta}}${}_2${{M}}${}_3${{Te}}${}_5$ ({{M}} = {{Ni}}, {{Pd}}) monolayers},\
  }\href {https://doi.org/10.1038/s41535-022-00498-8} {\bibfield  {journal}
  {\bibinfo  {journal} {npj Quantum Materials}\ }\textbf {\bibinfo {volume}
  {7}},\ \bibinfo {pages} {1} (\bibinfo {year} {2022})}\BibitemShut {NoStop}%
\bibitem [{\citenamefont {Reis}\ \emph {et~al.}(2017)\citenamefont {Reis},
  \citenamefont {Li}, \citenamefont {Dudy}, \citenamefont {Bauernfeind},
  \citenamefont {Glass}, \citenamefont {Hanke}, \citenamefont {Thomale},
  \citenamefont {Sch{\"a}fer},\ and\ \citenamefont
  {Claessen}}]{reis2017bismuthene}%
  \BibitemOpen
  \bibfield  {author} {\bibinfo {author} {\bibfnamefont {F.}~\bibnamefont
  {Reis}}, \bibinfo {author} {\bibfnamefont {G.}~\bibnamefont {Li}}, \bibinfo
  {author} {\bibfnamefont {L.}~\bibnamefont {Dudy}}, \bibinfo {author}
  {\bibfnamefont {M.}~\bibnamefont {Bauernfeind}}, \bibinfo {author}
  {\bibfnamefont {S.}~\bibnamefont {Glass}}, \bibinfo {author} {\bibfnamefont
  {W.}~\bibnamefont {Hanke}}, \bibinfo {author} {\bibfnamefont
  {R.}~\bibnamefont {Thomale}}, \bibinfo {author} {\bibfnamefont
  {J.}~\bibnamefont {Sch{\"a}fer}},\ and\ \bibinfo {author} {\bibfnamefont
  {R.}~\bibnamefont {Claessen}},\ }\bibfield  {title} {\bibinfo {title}
  {Bismuthene on a {{SiC}} substrate: {{A}} candidate for a high-temperature
  quantum spin {{Hall}} material},\ }\href
  {https://doi.org/10.1126/science.aai8142} {\bibfield  {journal} {\bibinfo
  {journal} {Science}\ }\textbf {\bibinfo {volume} {357}},\ \bibinfo {pages}
  {287} (\bibinfo {year} {2017})}\BibitemShut {NoStop}%
\bibitem [{\citenamefont {Chen}\ \emph {et~al.}(2021)\citenamefont {Chen},
  \citenamefont {Wu}, \citenamefont {Yu}, \citenamefont {Chen}, \citenamefont
  {Zhao}, \citenamefont {Sheng},\ and\ \citenamefont
  {Yang}}]{chen2021graphynea}%
  \BibitemOpen
  \bibfield  {author} {\bibinfo {author} {\bibfnamefont {C.}~\bibnamefont
  {Chen}}, \bibinfo {author} {\bibfnamefont {W.}~\bibnamefont {Wu}}, \bibinfo
  {author} {\bibfnamefont {Z.-M.}\ \bibnamefont {Yu}}, \bibinfo {author}
  {\bibfnamefont {Z.}~\bibnamefont {Chen}}, \bibinfo {author} {\bibfnamefont
  {Y.~X.}\ \bibnamefont {Zhao}}, \bibinfo {author} {\bibfnamefont {X.-L.}\
  \bibnamefont {Sheng}},\ and\ \bibinfo {author} {\bibfnamefont {S.~A.}\
  \bibnamefont {Yang}},\ }\bibfield  {title} {\bibinfo {title} {Graphyne as a
  second-order and real {{Chern}} topological insulator in two dimensions},\
  }\href {https://doi.org/10.1103/PhysRevB.104.085205} {\bibfield  {journal}
  {\bibinfo  {journal} {Physical Review B}\ }\textbf {\bibinfo {volume}
  {104}},\ \bibinfo {pages} {085205} (\bibinfo {year} {2021})}\BibitemShut
  {NoStop}%
\bibitem [{\citenamefont {Liu}\ \emph {et~al.}(2019)\citenamefont {Liu},
  \citenamefont {Zhao}, \citenamefont {Liu},\ and\ \citenamefont
  {Wang}}]{liu2019twodimensionalb}%
  \BibitemOpen
  \bibfield  {author} {\bibinfo {author} {\bibfnamefont {B.}~\bibnamefont
  {Liu}}, \bibinfo {author} {\bibfnamefont {G.}~\bibnamefont {Zhao}}, \bibinfo
  {author} {\bibfnamefont {Z.}~\bibnamefont {Liu}},\ and\ \bibinfo {author}
  {\bibfnamefont {Z.~F.}\ \bibnamefont {Wang}},\ }\bibfield  {title} {\bibinfo
  {title} {Two-{{Dimensional Quadrupole Topological Insulator}} in
  {$\gamma$}-{{Graphyne}}},\ }\href
  {https://doi.org/10.1021/acs.nanolett.9b02719} {\bibfield  {journal}
  {\bibinfo  {journal} {Nano Letters}\ }\textbf {\bibinfo {volume} {19}},\
  \bibinfo {pages} {6492} (\bibinfo {year} {2019})}\BibitemShut {NoStop}%
\bibitem [{\citenamefont {Park}\ \emph {et~al.}(2019)\citenamefont {Park},
  \citenamefont {Kim}, \citenamefont {Cho},\ and\ \citenamefont
  {Lee}}]{park2019higherordera}%
  \BibitemOpen
  \bibfield  {author} {\bibinfo {author} {\bibfnamefont {M.~J.}\ \bibnamefont
  {Park}}, \bibinfo {author} {\bibfnamefont {Y.}~\bibnamefont {Kim}}, \bibinfo
  {author} {\bibfnamefont {G.~Y.}\ \bibnamefont {Cho}},\ and\ \bibinfo {author}
  {\bibfnamefont {S.}~\bibnamefont {Lee}},\ }\bibfield  {title} {\bibinfo
  {title} {Higher-{{Order Topological Insulator}} in {{Twisted Bilayer
  Graphene}}},\ }\href {https://doi.org/10.1103/PhysRevLett.123.216803}
  {\bibfield  {journal} {\bibinfo  {journal} {Physical Review Letters}\
  }\textbf {\bibinfo {volume} {123}},\ \bibinfo {pages} {216803} (\bibinfo
  {year} {2019})}\BibitemShut {NoStop}%
\bibitem [{\citenamefont {Xue}\ \emph {et~al.}(2021)\citenamefont {Xue},
  \citenamefont {Huan}, \citenamefont {Zhao}, \citenamefont {Luo},
  \citenamefont {Zhang},\ and\ \citenamefont {Yang}}]{xue2021higherordera}%
  \BibitemOpen
  \bibfield  {author} {\bibinfo {author} {\bibfnamefont {Y.}~\bibnamefont
  {Xue}}, \bibinfo {author} {\bibfnamefont {H.}~\bibnamefont {Huan}}, \bibinfo
  {author} {\bibfnamefont {B.}~\bibnamefont {Zhao}}, \bibinfo {author}
  {\bibfnamefont {Y.}~\bibnamefont {Luo}}, \bibinfo {author} {\bibfnamefont
  {Z.}~\bibnamefont {Zhang}},\ and\ \bibinfo {author} {\bibfnamefont
  {Z.}~\bibnamefont {Yang}},\ }\bibfield  {title} {\bibinfo {title}
  {Higher-order topological insulators in two-dimensional {{Dirac}}
  materials},\ }\href {https://doi.org/10.1103/PhysRevResearch.3.L042044}
  {\bibfield  {journal} {\bibinfo  {journal} {Physical Review Research}\
  }\textbf {\bibinfo {volume} {3}},\ \bibinfo {pages} {L042044} (\bibinfo
  {year} {2021})}\BibitemShut {NoStop}%
\bibitem [{\citenamefont {Takahashi}\ \emph {et~al.}(2021)\citenamefont
  {Takahashi}, \citenamefont {Zhang},\ and\ \citenamefont
  {Murakami}}]{takahashi2021general}%
  \BibitemOpen
  \bibfield  {author} {\bibinfo {author} {\bibfnamefont {R.}~\bibnamefont
  {Takahashi}}, \bibinfo {author} {\bibfnamefont {T.}~\bibnamefont {Zhang}},\
  and\ \bibinfo {author} {\bibfnamefont {S.}~\bibnamefont {Murakami}},\
  }\bibfield  {title} {\bibinfo {title} {General corner charge formula in
  two-dimensional {{C}}${}_n$ $-$symmetric higher-order topological
  insulators},\ }\href {https://doi.org/10.1103/PhysRevB.103.205123} {\bibfield
   {journal} {\bibinfo  {journal} {Physical Review B}\ }\textbf {\bibinfo
  {volume} {103}},\ \bibinfo {pages} {205123} (\bibinfo {year}
  {2021})}\BibitemShut {NoStop}%
\bibitem [{\citenamefont {Qian}\ \emph {et~al.}(2022)\citenamefont {Qian},
  \citenamefont {Liu}, \citenamefont {Liu},\ and\ \citenamefont
  {Yao}}]{qian2022c}%
  \BibitemOpen
  \bibfield  {author} {\bibinfo {author} {\bibfnamefont {S.}~\bibnamefont
  {Qian}}, \bibinfo {author} {\bibfnamefont {G.-B.}\ \bibnamefont {Liu}},
  \bibinfo {author} {\bibfnamefont {C.-C.}\ \bibnamefont {Liu}},\ and\ \bibinfo
  {author} {\bibfnamefont {Y.}~\bibnamefont {Yao}},\ }\bibfield  {title}
  {\bibinfo {title} {{{C}}${}_n$ -symmetric higher-order topological
  crystalline insulators in atomically thin transition metal dichalcogenides},\
  }\href {https://doi.org/10.1103/PhysRevB.105.045417} {\bibfield  {journal}
  {\bibinfo  {journal} {Physical Review B}\ }\textbf {\bibinfo {volume}
  {105}},\ \bibinfo {pages} {045417} (\bibinfo {year} {2022})}\BibitemShut
  {NoStop}%
\bibitem [{\citenamefont {Zeng}\ \emph {et~al.}(2021)\citenamefont {Zeng},
  \citenamefont {Liu}, \citenamefont {Jiang}, \citenamefont {Sun},\ and\
  \citenamefont {Xie}}]{zeng2021multiorbitala}%
  \BibitemOpen
  \bibfield  {author} {\bibinfo {author} {\bibfnamefont {J.}~\bibnamefont
  {Zeng}}, \bibinfo {author} {\bibfnamefont {H.}~\bibnamefont {Liu}}, \bibinfo
  {author} {\bibfnamefont {H.}~\bibnamefont {Jiang}}, \bibinfo {author}
  {\bibfnamefont {Q.-F.}\ \bibnamefont {Sun}},\ and\ \bibinfo {author}
  {\bibfnamefont {X.~C.}\ \bibnamefont {Xie}},\ }\bibfield  {title} {\bibinfo
  {title} {Multiorbital model reveals a second-order topological insulator in
  {{1H}} transition metal dichalcogenides},\ }\href
  {https://doi.org/10.1103/PhysRevB.104.L161108} {\bibfield  {journal}
  {\bibinfo  {journal} {Physical Review B}\ }\textbf {\bibinfo {volume}
  {104}},\ \bibinfo {pages} {L161108} (\bibinfo {year} {2021})}\BibitemShut
  {NoStop}%
\bibitem [{\citenamefont {Li}\ \emph {et~al.}(2022{\natexlab{a}})\citenamefont
  {Li}, \citenamefont {Zhou}, \citenamefont {Yan}, \citenamefont {Peng},
  \citenamefont {Ma},\ and\ \citenamefont {Sun}}]{li2022secondorder}%
  \BibitemOpen
  \bibfield  {author} {\bibinfo {author} {\bibfnamefont {Z.}~\bibnamefont
  {Li}}, \bibinfo {author} {\bibfnamefont {P.}~\bibnamefont {Zhou}}, \bibinfo
  {author} {\bibfnamefont {Q.}~\bibnamefont {Yan}}, \bibinfo {author}
  {\bibfnamefont {X.}~\bibnamefont {Peng}}, \bibinfo {author} {\bibfnamefont
  {Z.}~\bibnamefont {Ma}},\ and\ \bibinfo {author} {\bibfnamefont
  {L.}~\bibnamefont {Sun}},\ }\bibfield  {title} {\bibinfo {title}
  {Second-order topological insulator in two-dimensional {{C}}${}_2${{N}} and
  its derivatives},\ }\href {https://doi.org/10.1103/PhysRevB.106.085126}
  {\bibfield  {journal} {\bibinfo  {journal} {Physical Review B}\ }\textbf
  {\bibinfo {volume} {106}},\ \bibinfo {pages} {085126} (\bibinfo {year}
  {2022}{\natexlab{a}})}\BibitemShut {NoStop}%
\bibitem [{\citenamefont {Qian}\ \emph {et~al.}(2021)\citenamefont {Qian},
  \citenamefont {Liu},\ and\ \citenamefont {Yao}}]{qian2021secondorderb}%
  \BibitemOpen
  \bibfield  {author} {\bibinfo {author} {\bibfnamefont {S.}~\bibnamefont
  {Qian}}, \bibinfo {author} {\bibfnamefont {C.-C.}\ \bibnamefont {Liu}},\ and\
  \bibinfo {author} {\bibfnamefont {Y.}~\bibnamefont {Yao}},\ }\bibfield
  {title} {\bibinfo {title} {Second-order topological insulator state in
  hexagonal lattices and its abundant material candidates},\ }\href
  {https://doi.org/10.1103/PhysRevB.104.245427} {\bibfield  {journal} {\bibinfo
   {journal} {Physical Review B}\ }\textbf {\bibinfo {volume} {104}},\ \bibinfo
  {pages} {245427} (\bibinfo {year} {2021})}\BibitemShut {NoStop}%
\bibitem [{\citenamefont {Li}\ \emph {et~al.}(2023)\citenamefont {Li},
  \citenamefont {Mao}, \citenamefont {Wu}, \citenamefont {Huang}, \citenamefont
  {Dai},\ and\ \citenamefont {Niu}}]{li2023robusta}%
  \BibitemOpen
  \bibfield  {author} {\bibinfo {author} {\bibfnamefont {R.}~\bibnamefont
  {Li}}, \bibinfo {author} {\bibfnamefont {N.}~\bibnamefont {Mao}}, \bibinfo
  {author} {\bibfnamefont {X.}~\bibnamefont {Wu}}, \bibinfo {author}
  {\bibfnamefont {B.}~\bibnamefont {Huang}}, \bibinfo {author} {\bibfnamefont
  {Y.}~\bibnamefont {Dai}},\ and\ \bibinfo {author} {\bibfnamefont
  {C.}~\bibnamefont {Niu}},\ }\bibfield  {title} {\bibinfo {title} {Robust
  {{Second-Order Topological Insulators}} with {{Giant Valley Polarization}} in
  {{Two-Dimensional Honeycomb Ferromagnets}}},\ }\href
  {https://doi.org/10.1021/acs.nanolett.2c03680} {\bibfield  {journal}
  {\bibinfo  {journal} {Nano Letters}\ }\textbf {\bibinfo {volume} {23}},\
  \bibinfo {pages} {91} (\bibinfo {year} {2023})}\BibitemShut {NoStop}%
\bibitem [{\citenamefont {Costa}\ \emph {et~al.}(2021)\citenamefont {Costa},
  \citenamefont {Schleder}, \citenamefont {Mera~Acosta}, \citenamefont
  {Padilha}, \citenamefont {Cerasoli}, \citenamefont {Buongiorno~Nardelli},\
  and\ \citenamefont {Fazzio}}]{costa2021discovery}%
  \BibitemOpen
  \bibfield  {author} {\bibinfo {author} {\bibfnamefont {M.}~\bibnamefont
  {Costa}}, \bibinfo {author} {\bibfnamefont {G.~R.}\ \bibnamefont {Schleder}},
  \bibinfo {author} {\bibfnamefont {C.}~\bibnamefont {Mera~Acosta}}, \bibinfo
  {author} {\bibfnamefont {A.~C.~M.}\ \bibnamefont {Padilha}}, \bibinfo
  {author} {\bibfnamefont {F.}~\bibnamefont {Cerasoli}}, \bibinfo {author}
  {\bibfnamefont {M.}~\bibnamefont {Buongiorno~Nardelli}},\ and\ \bibinfo
  {author} {\bibfnamefont {A.}~\bibnamefont {Fazzio}},\ }\bibfield  {title}
  {\bibinfo {title} {Discovery of higher-order topological insulators using the
  spin {{Hall}} conductivity as a topology signature},\ }\href
  {https://doi.org/10.1038/s41524-021-00518-4} {\bibfield  {journal} {\bibinfo
  {journal} {npj Computational Materials}\ }\textbf {\bibinfo {volume} {7}},\
  \bibinfo {pages} {1} (\bibinfo {year} {2021})}\BibitemShut {NoStop}%
\bibitem [{\citenamefont {Liu}\ \emph {et~al.}(2022)\citenamefont {Liu},
  \citenamefont {Ren}, \citenamefont {Han}, \citenamefont {Niu},\ and\
  \citenamefont {Qiao}}]{liu2022secondorder}%
  \BibitemOpen
  \bibfield  {author} {\bibinfo {author} {\bibfnamefont {Z.}~\bibnamefont
  {Liu}}, \bibinfo {author} {\bibfnamefont {Y.}~\bibnamefont {Ren}}, \bibinfo
  {author} {\bibfnamefont {Y.}~\bibnamefont {Han}}, \bibinfo {author}
  {\bibfnamefont {Q.}~\bibnamefont {Niu}},\ and\ \bibinfo {author}
  {\bibfnamefont {Z.}~\bibnamefont {Qiao}},\ }\bibfield  {title} {\bibinfo
  {title} {Second-order topological insulator in van der {{Waals}}
  heterostructures of
  {{CoBr}}${}_2$/{{Pt}}${}_2${{HgSe}}${}_3$/{{CoBr}}${}_2$},\ }\href
  {https://doi.org/10.1103/PhysRevB.106.195303} {\bibfield  {journal} {\bibinfo
   {journal} {Physical Review B}\ }\textbf {\bibinfo {volume} {106}},\ \bibinfo
  {pages} {195303} (\bibinfo {year} {2022})}\BibitemShut {NoStop}%
\bibitem [{\citenamefont {Chen}\ \emph {et~al.}(2020)\citenamefont {Chen},
  \citenamefont {Song}, \citenamefont {Zhao}, \citenamefont {Chen},
  \citenamefont {Yu}, \citenamefont {Sheng},\ and\ \citenamefont
  {Yang}}]{chen2020universala}%
  \BibitemOpen
  \bibfield  {author} {\bibinfo {author} {\bibfnamefont {C.}~\bibnamefont
  {Chen}}, \bibinfo {author} {\bibfnamefont {Z.}~\bibnamefont {Song}}, \bibinfo
  {author} {\bibfnamefont {J.-Z.}\ \bibnamefont {Zhao}}, \bibinfo {author}
  {\bibfnamefont {Z.}~\bibnamefont {Chen}}, \bibinfo {author} {\bibfnamefont
  {Z.-M.}\ \bibnamefont {Yu}}, \bibinfo {author} {\bibfnamefont {X.-L.}\
  \bibnamefont {Sheng}},\ and\ \bibinfo {author} {\bibfnamefont {S.~A.}\
  \bibnamefont {Yang}},\ }\bibfield  {title} {\bibinfo {title} {Universal
  {{Approach}} to {{Magnetic Second-Order Topological Insulator}}},\ }\href
  {https://doi.org/10.1103/PhysRevLett.125.056402} {\bibfield  {journal}
  {\bibinfo  {journal} {Physical Review Letters}\ }\textbf {\bibinfo {volume}
  {125}},\ \bibinfo {pages} {056402} (\bibinfo {year} {2020})}\BibitemShut
  {NoStop}%
\bibitem [{\citenamefont {Ren}\ \emph {et~al.}(2020)\citenamefont {Ren},
  \citenamefont {Qiao},\ and\ \citenamefont {Niu}}]{ren2020engineering}%
  \BibitemOpen
  \bibfield  {author} {\bibinfo {author} {\bibfnamefont {Y.}~\bibnamefont
  {Ren}}, \bibinfo {author} {\bibfnamefont {Z.}~\bibnamefont {Qiao}},\ and\
  \bibinfo {author} {\bibfnamefont {Q.}~\bibnamefont {Niu}},\ }\bibfield
  {title} {\bibinfo {title} {Engineering {{Corner States}} from
  {{Two-Dimensional Topological Insulators}}},\ }\href
  {https://doi.org/10.1103/PhysRevLett.124.166804} {\bibfield  {journal}
  {\bibinfo  {journal} {Physical Review Letters}\ }\textbf {\bibinfo {volume}
  {124}},\ \bibinfo {pages} {166804} (\bibinfo {year} {2020})}\BibitemShut
  {NoStop}%
\bibitem [{\citenamefont {Mu}\ \emph {et~al.}(2022)\citenamefont {Mu},
  \citenamefont {Zhao}, \citenamefont {Zhang},\ and\ \citenamefont
  {Wang}}]{RN2831}%
  \BibitemOpen
  \bibfield  {author} {\bibinfo {author} {\bibfnamefont {H.}~\bibnamefont
  {Mu}}, \bibinfo {author} {\bibfnamefont {G.}~\bibnamefont {Zhao}}, \bibinfo
  {author} {\bibfnamefont {H.}~\bibnamefont {Zhang}},\ and\ \bibinfo {author}
  {\bibfnamefont {Z.}~\bibnamefont {Wang}},\ }\bibfield  {title} {\bibinfo
  {title} {Antiferromagnetic second-order topological insulator with fractional
  mass-kink},\ }\bibfield  {journal} {\bibinfo  {journal} {npj Computational
  Materials}\ }\textbf {\bibinfo {volume} {8}},\ \href
  {https://doi.org/10.1038/s41524-022-00761-3} {10.1038/s41524-022-00761-3}
  (\bibinfo {year} {2022})\BibitemShut {NoStop}%
\bibitem [{\citenamefont {Mao}\ \emph {et~al.}(2023)\citenamefont {Mao},
  \citenamefont {Li}, \citenamefont {Zou}, \citenamefont {Dai}, \citenamefont
  {Huang},\ and\ \citenamefont {Niu}}]{mao2023ferroelectric}%
  \BibitemOpen
  \bibfield  {author} {\bibinfo {author} {\bibfnamefont {N.}~\bibnamefont
  {Mao}}, \bibinfo {author} {\bibfnamefont {R.}~\bibnamefont {Li}}, \bibinfo
  {author} {\bibfnamefont {X.}~\bibnamefont {Zou}}, \bibinfo {author}
  {\bibfnamefont {Y.}~\bibnamefont {Dai}}, \bibinfo {author} {\bibfnamefont
  {B.}~\bibnamefont {Huang}},\ and\ \bibinfo {author} {\bibfnamefont
  {C.}~\bibnamefont {Niu}},\ }\bibfield  {title} {\bibinfo {title}
  {Ferroelectric higher-order topological insulator in two dimensions},\ }\href
  {https://doi.org/10.1103/PhysRevB.107.045125} {\bibfield  {journal} {\bibinfo
   {journal} {Physical Review B}\ }\textbf {\bibinfo {volume} {107}},\ \bibinfo
  {pages} {045125} (\bibinfo {year} {2023})}\BibitemShut {NoStop}%
\bibitem [{\citenamefont {Zhang}\ \emph
  {et~al.}(2023{\natexlab{a}})\citenamefont {Zhang}, \citenamefont {He},
  \citenamefont {Liu}, \citenamefont {Dai}, \citenamefont {Liu}, \citenamefont
  {Chen}, \citenamefont {Wu}, \citenamefont {Zhu},\ and\ \citenamefont
  {Yang}}]{zhang2023magnetic}%
  \BibitemOpen
  \bibfield  {author} {\bibinfo {author} {\bibfnamefont {X.}~\bibnamefont
  {Zhang}}, \bibinfo {author} {\bibfnamefont {T.}~\bibnamefont {He}}, \bibinfo
  {author} {\bibfnamefont {Y.}~\bibnamefont {Liu}}, \bibinfo {author}
  {\bibfnamefont {X.}~\bibnamefont {Dai}}, \bibinfo {author} {\bibfnamefont
  {G.}~\bibnamefont {Liu}}, \bibinfo {author} {\bibfnamefont {C.}~\bibnamefont
  {Chen}}, \bibinfo {author} {\bibfnamefont {W.}~\bibnamefont {Wu}}, \bibinfo
  {author} {\bibfnamefont {J.}~\bibnamefont {Zhu}},\ and\ \bibinfo {author}
  {\bibfnamefont {S.~A.}\ \bibnamefont {Yang}},\ }\href@noop {} {\bibinfo
  {title} {Magnetic real chern insulator in 2d metal-organic frameworks}}
  (\bibinfo {year} {2023}{\natexlab{a}}),\ \Eprint
  {https://arxiv.org/abs/2303.09218} {arXiv:2303.09218} \BibitemShut {NoStop}%
\bibitem [{\citenamefont {Gong}\ \emph {et~al.}(2024)\citenamefont {Gong},
  \citenamefont {Wang}, \citenamefont {Han}, \citenamefont {Cheng},
  \citenamefont {Wang}, \citenamefont {Yu},\ and\ \citenamefont
  {Yao}}]{adma.202402232}%
  \BibitemOpen
  \bibfield  {author} {\bibinfo {author} {\bibfnamefont {J.}~\bibnamefont
  {Gong}}, \bibinfo {author} {\bibfnamefont {Y.}~\bibnamefont {Wang}}, \bibinfo
  {author} {\bibfnamefont {Y.}~\bibnamefont {Han}}, \bibinfo {author}
  {\bibfnamefont {Z.}~\bibnamefont {Cheng}}, \bibinfo {author} {\bibfnamefont
  {X.}~\bibnamefont {Wang}}, \bibinfo {author} {\bibfnamefont {Z.-M.}\
  \bibnamefont {Yu}},\ and\ \bibinfo {author} {\bibfnamefont {Y.}~\bibnamefont
  {Yao}},\ }\bibfield  {title} {\bibinfo {title} {Hidden real topology and
  unusual magnetoelectric responses in two-dimensional antiferromagnets},\
  }\href {https://doi.org/https://doi.org/10.1002/adma.202402232} {\bibfield
  {journal} {\bibinfo  {journal} {Advanced Materials}\ ,\ \bibinfo {pages}
  {2402232}} (\bibinfo {year} {2024})}\BibitemShut {NoStop}%
\bibitem [{\citenamefont {Pahomi}\ \emph {et~al.}(2020)\citenamefont {Pahomi},
  \citenamefont {Sigrist},\ and\ \citenamefont
  {Soluyanov}}]{pahomi2020braiding}%
  \BibitemOpen
  \bibfield  {author} {\bibinfo {author} {\bibfnamefont {T.~E.}\ \bibnamefont
  {Pahomi}}, \bibinfo {author} {\bibfnamefont {M.}~\bibnamefont {Sigrist}},\
  and\ \bibinfo {author} {\bibfnamefont {A.~A.}\ \bibnamefont {Soluyanov}},\
  }\bibfield  {title} {\bibinfo {title} {Braiding {{Majorana}} corner modes in
  a second-order topological superconductor},\ }\href
  {https://doi.org/10.1103/PhysRevResearch.2.032068} {\bibfield  {journal}
  {\bibinfo  {journal} {Physical Review Research}\ }\textbf {\bibinfo {volume}
  {2}},\ \bibinfo {pages} {032068} (\bibinfo {year} {2020})}\BibitemShut
  {NoStop}%
\bibitem [{\citenamefont {Zhang}\ \emph {et~al.}(2020)\citenamefont {Zhang},
  \citenamefont {Rui}, \citenamefont {Calzona}, \citenamefont {Choi},
  \citenamefont {Schnyder},\ and\ \citenamefont
  {Trauzettel}}]{zhang2020topological}%
  \BibitemOpen
  \bibfield  {author} {\bibinfo {author} {\bibfnamefont {S.-B.}\ \bibnamefont
  {Zhang}}, \bibinfo {author} {\bibfnamefont {W.~B.}\ \bibnamefont {Rui}},
  \bibinfo {author} {\bibfnamefont {A.}~\bibnamefont {Calzona}}, \bibinfo
  {author} {\bibfnamefont {S.-J.}\ \bibnamefont {Choi}}, \bibinfo {author}
  {\bibfnamefont {A.~P.}\ \bibnamefont {Schnyder}},\ and\ \bibinfo {author}
  {\bibfnamefont {B.}~\bibnamefont {Trauzettel}},\ }\bibfield  {title}
  {\bibinfo {title} {Topological and holonomic quantum computation based on
  second-order topological superconductors},\ }\href
  {https://doi.org/10.1103/PhysRevResearch.2.043025} {\bibfield  {journal}
  {\bibinfo  {journal} {Physical Review Research}\ }\textbf {\bibinfo {volume}
  {2}},\ \bibinfo {pages} {043025} (\bibinfo {year} {2020})}\BibitemShut
  {NoStop}%
\bibitem [{\citenamefont {Li}\ \emph {et~al.}(2022{\natexlab{b}})\citenamefont
  {Li}, \citenamefont {Geier}, \citenamefont {Ingham},\ and\ \citenamefont
  {D~Scammell}}]{li2022higherorder}%
  \BibitemOpen
  \bibfield  {author} {\bibinfo {author} {\bibfnamefont {T.}~\bibnamefont
  {Li}}, \bibinfo {author} {\bibfnamefont {M.}~\bibnamefont {Geier}}, \bibinfo
  {author} {\bibfnamefont {J.}~\bibnamefont {Ingham}},\ and\ \bibinfo {author}
  {\bibfnamefont {H.}~\bibnamefont {D~Scammell}},\ }\bibfield  {title}
  {\bibinfo {title} {Higher-order topological superconductivity from repulsive
  interactions in kagome and honeycomb systems},\ }\href
  {https://doi.org/10.1088/2053-1583/ac4060} {\bibfield  {journal} {\bibinfo
  {journal} {2D Materials}\ }\textbf {\bibinfo {volume} {9}},\ \bibinfo {pages}
  {015031} (\bibinfo {year} {2022}{\natexlab{b}})}\BibitemShut {NoStop}%
\bibitem [{\citenamefont {Ameta}\ \emph {et~al.}(2023)\citenamefont {Ameta},
  \citenamefont {Bhatt},\ and\ \citenamefont {Ameta}}]{ameta2023quantum}%
  \BibitemOpen
  \bibinfo {editor} {\bibfnamefont {R.}~\bibnamefont {Ameta}}, \bibinfo
  {editor} {\bibfnamefont {J.~P.}\ \bibnamefont {Bhatt}},\ and\ \bibinfo
  {editor} {\bibfnamefont {S.~C.}\ \bibnamefont {Ameta}},\ eds.,\ \href@noop {}
  {\emph {\bibinfo {title} {Quantum Dots: Fundamentals, Synthesis and
  Applications}}}\ (\bibinfo  {publisher} {Elsevier},\ \bibinfo {address}
  {Amsterdam, Netherlands ; Cambridge, MA},\ \bibinfo {year}
  {2023})\BibitemShut {NoStop}%
\bibitem [{\citenamefont {Jung}\ \emph {et~al.}(2021)\citenamefont {Jung},
  \citenamefont {Ahn},\ and\ \citenamefont {Klimov}}]{jung2021prospectsb}%
  \BibitemOpen
  \bibfield  {author} {\bibinfo {author} {\bibfnamefont {H.}~\bibnamefont
  {Jung}}, \bibinfo {author} {\bibfnamefont {N.}~\bibnamefont {Ahn}},\ and\
  \bibinfo {author} {\bibfnamefont {V.~I.}\ \bibnamefont {Klimov}},\ }\bibfield
   {title} {\bibinfo {title} {Prospects and challenges of colloidal quantum dot
  laser diodes},\ }\href {https://doi.org/10.1038/s41566-021-00827-6}
  {\bibfield  {journal} {\bibinfo  {journal} {Nature Photonics}\ }\textbf
  {\bibinfo {volume} {15}},\ \bibinfo {pages} {643} (\bibinfo {year}
  {2021})}\BibitemShut {NoStop}%
\bibitem [{\citenamefont {Medintz}\ \emph {et~al.}(2005)\citenamefont
  {Medintz}, \citenamefont {Uyeda}, \citenamefont {Goldman},\ and\
  \citenamefont {Mattoussi}}]{medintz2005medintz}%
  \BibitemOpen
  \bibfield  {author} {\bibinfo {author} {\bibfnamefont {I.}~\bibnamefont
  {Medintz}}, \bibinfo {author} {\bibfnamefont {H.}~\bibnamefont {Uyeda}},
  \bibinfo {author} {\bibfnamefont {E.}~\bibnamefont {Goldman}},\ and\ \bibinfo
  {author} {\bibfnamefont {H.}~\bibnamefont {Mattoussi}},\ }\bibfield  {title}
  {\bibinfo {title} {Medintz, {{I}}.{{L}}., {{Uyeda}}, {{H}}.{{T}}.,
  {{Goldman}}, {{E}}.{{R}}. \& {{Mattoussi}}, {{H}}. {{Quantum}} dot
  bioconjugates for imaging, labelling and sensing. {{Nat}}. {{Mater}}. 4,
  435-446},\ }\href {https://doi.org/10.1038/nmat1390} {\bibfield  {journal}
  {\bibinfo  {journal} {Nature materials}\ }\textbf {\bibinfo {volume} {4}},\
  \bibinfo {pages} {435} (\bibinfo {year} {2005})}\BibitemShut {NoStop}%
\bibitem [{\citenamefont {Ying~Lim}\ \emph {et~al.}(2015)\citenamefont
  {Ying~Lim}, \citenamefont {Shen},\ and\ \citenamefont
  {Gao}}]{yinglim2015carbon}%
  \BibitemOpen
  \bibfield  {author} {\bibinfo {author} {\bibfnamefont {S.}~\bibnamefont
  {Ying~Lim}}, \bibinfo {author} {\bibfnamefont {W.}~\bibnamefont {Shen}},\
  and\ \bibinfo {author} {\bibfnamefont {Z.}~\bibnamefont {Gao}},\ }\bibfield
  {title} {\bibinfo {title} {Carbon quantum dots and their applications},\
  }\href {https://doi.org/10.1039/C4CS00269E} {\bibfield  {journal} {\bibinfo
  {journal} {Chemical Society Reviews}\ }\textbf {\bibinfo {volume} {44}},\
  \bibinfo {pages} {362} (\bibinfo {year} {2015})}\BibitemShut {NoStop}%
\bibitem [{\citenamefont {Bera}\ \emph {et~al.}(2010)\citenamefont {Bera},
  \citenamefont {Qian}, \citenamefont {Tseng},\ and\ \citenamefont
  {Holloway}}]{bera2010quantum}%
  \BibitemOpen
  \bibfield  {author} {\bibinfo {author} {\bibfnamefont {D.}~\bibnamefont
  {Bera}}, \bibinfo {author} {\bibfnamefont {L.}~\bibnamefont {Qian}}, \bibinfo
  {author} {\bibfnamefont {T.-K.}\ \bibnamefont {Tseng}},\ and\ \bibinfo
  {author} {\bibfnamefont {P.~H.}\ \bibnamefont {Holloway}},\ }\bibfield
  {title} {\bibinfo {title} {Quantum {{Dots}} and {{Their Multimodal
  Applications}}: {{A Review}}},\ }\href {https://doi.org/10.3390/ma3042260}
  {\bibfield  {journal} {\bibinfo  {journal} {Materials}\ }\textbf {\bibinfo
  {volume} {3}},\ \bibinfo {pages} {2260} (\bibinfo {year} {2010})}\BibitemShut
  {NoStop}%
\bibitem [{\citenamefont {Eich}\ \emph {et~al.}(2018)\citenamefont {Eich},
  \citenamefont {Herman}, \citenamefont {Pisoni}, \citenamefont {Overweg},
  \citenamefont {Kurzmann}, \citenamefont {Lee}, \citenamefont {Rickhaus},
  \citenamefont {Watanabe}, \citenamefont {Taniguchi}, \citenamefont {Sigrist},
  \citenamefont {Ihn},\ and\ \citenamefont {Ensslin}}]{eich2018spin}%
  \BibitemOpen
  \bibfield  {author} {\bibinfo {author} {\bibfnamefont {M.}~\bibnamefont
  {Eich}}, \bibinfo {author} {\bibfnamefont {F.}~\bibnamefont {Herman}},
  \bibinfo {author} {\bibfnamefont {R.}~\bibnamefont {Pisoni}}, \bibinfo
  {author} {\bibfnamefont {H.}~\bibnamefont {Overweg}}, \bibinfo {author}
  {\bibfnamefont {A.}~\bibnamefont {Kurzmann}}, \bibinfo {author}
  {\bibfnamefont {Y.}~\bibnamefont {Lee}}, \bibinfo {author} {\bibfnamefont
  {P.}~\bibnamefont {Rickhaus}}, \bibinfo {author} {\bibfnamefont
  {K.}~\bibnamefont {Watanabe}}, \bibinfo {author} {\bibfnamefont
  {T.}~\bibnamefont {Taniguchi}}, \bibinfo {author} {\bibfnamefont
  {M.}~\bibnamefont {Sigrist}}, \bibinfo {author} {\bibfnamefont
  {T.}~\bibnamefont {Ihn}},\ and\ \bibinfo {author} {\bibfnamefont
  {K.}~\bibnamefont {Ensslin}},\ }\bibfield  {title} {\bibinfo {title} {Spin
  and {{Valley States}} in {{Gate-Defined Bilayer Graphene Quantum Dots}}},\
  }\href {https://doi.org/10.1103/PhysRevX.8.031023} {\bibfield  {journal}
  {\bibinfo  {journal} {Physical Review X}\ }\textbf {\bibinfo {volume} {8}},\
  \bibinfo {pages} {031023} (\bibinfo {year} {2018})}\BibitemShut {NoStop}%
\bibitem [{\citenamefont {Kurzmann}\ \emph {et~al.}(2019)\citenamefont
  {Kurzmann}, \citenamefont {Overweg}, \citenamefont {Eich}, \citenamefont
  {Pally}, \citenamefont {Rickhaus}, \citenamefont {Pisoni}, \citenamefont
  {Lee}, \citenamefont {Watanabe}, \citenamefont {Taniguchi}, \citenamefont
  {Ihn},\ and\ \citenamefont {Ensslin}}]{kurzmann2019charge}%
  \BibitemOpen
  \bibfield  {author} {\bibinfo {author} {\bibfnamefont {A.}~\bibnamefont
  {Kurzmann}}, \bibinfo {author} {\bibfnamefont {H.}~\bibnamefont {Overweg}},
  \bibinfo {author} {\bibfnamefont {M.}~\bibnamefont {Eich}}, \bibinfo {author}
  {\bibfnamefont {A.}~\bibnamefont {Pally}}, \bibinfo {author} {\bibfnamefont
  {P.}~\bibnamefont {Rickhaus}}, \bibinfo {author} {\bibfnamefont
  {R.}~\bibnamefont {Pisoni}}, \bibinfo {author} {\bibfnamefont
  {Y.}~\bibnamefont {Lee}}, \bibinfo {author} {\bibfnamefont {K.}~\bibnamefont
  {Watanabe}}, \bibinfo {author} {\bibfnamefont {T.}~\bibnamefont {Taniguchi}},
  \bibinfo {author} {\bibfnamefont {T.}~\bibnamefont {Ihn}},\ and\ \bibinfo
  {author} {\bibfnamefont {K.}~\bibnamefont {Ensslin}},\ }\bibfield  {title}
  {\bibinfo {title} {Charge {{Detection}} in {{Gate-Defined Bilayer Graphene
  Quantum Dots}}},\ }\href {https://doi.org/10.1021/acs.nanolett.9b01617}
  {\bibfield  {journal} {\bibinfo  {journal} {Nano Letters}\ }\textbf {\bibinfo
  {volume} {19}},\ \bibinfo {pages} {5216} (\bibinfo {year}
  {2019})}\BibitemShut {NoStop}%
\bibitem [{\citenamefont {Banszerus}\ \emph {et~al.}(2020)\citenamefont
  {Banszerus}, \citenamefont {Rothstein}, \citenamefont {Fabian}, \citenamefont
  {M{\"o}ller}, \citenamefont {Icking}, \citenamefont {Trellenkamp},
  \citenamefont {Lentz}, \citenamefont {Neumaier}, \citenamefont {Watanabe},
  \citenamefont {Taniguchi}, \citenamefont {Libisch}, \citenamefont {Volk},\
  and\ \citenamefont {Stampfer}}]{banszerus2020electron}%
  \BibitemOpen
  \bibfield  {author} {\bibinfo {author} {\bibfnamefont {L.}~\bibnamefont
  {Banszerus}}, \bibinfo {author} {\bibfnamefont {A.}~\bibnamefont
  {Rothstein}}, \bibinfo {author} {\bibfnamefont {T.}~\bibnamefont {Fabian}},
  \bibinfo {author} {\bibfnamefont {S.}~\bibnamefont {M{\"o}ller}}, \bibinfo
  {author} {\bibfnamefont {E.}~\bibnamefont {Icking}}, \bibinfo {author}
  {\bibfnamefont {S.}~\bibnamefont {Trellenkamp}}, \bibinfo {author}
  {\bibfnamefont {F.}~\bibnamefont {Lentz}}, \bibinfo {author} {\bibfnamefont
  {D.}~\bibnamefont {Neumaier}}, \bibinfo {author} {\bibfnamefont
  {K.}~\bibnamefont {Watanabe}}, \bibinfo {author} {\bibfnamefont
  {T.}~\bibnamefont {Taniguchi}}, \bibinfo {author} {\bibfnamefont
  {F.}~\bibnamefont {Libisch}}, \bibinfo {author} {\bibfnamefont
  {C.}~\bibnamefont {Volk}},\ and\ \bibinfo {author} {\bibfnamefont
  {C.}~\bibnamefont {Stampfer}},\ }\bibfield  {title} {\bibinfo {title}
  {Electron--{{Hole Crossover}} in {{Gate-Controlled Bilayer Graphene Quantum
  Dots}}},\ }\href {https://doi.org/10.1021/acs.nanolett.0c03227} {\bibfield
  {journal} {\bibinfo  {journal} {Nano Letters}\ }\textbf {\bibinfo {volume}
  {20}},\ \bibinfo {pages} {7709} (\bibinfo {year} {2020})}\BibitemShut
  {NoStop}%
\bibitem [{\citenamefont {Moreels}\ \emph {et~al.}(2009)\citenamefont
  {Moreels}, \citenamefont {Lambert}, \citenamefont {Smeets}, \citenamefont
  {De~Muynck}, \citenamefont {Nollet}, \citenamefont {Martins}, \citenamefont
  {Vanhaecke}, \citenamefont {Vantomme}, \citenamefont {Delerue}, \citenamefont
  {Allan},\ and\ \citenamefont {Hens}}]{moreels2009sizedependent}%
  \BibitemOpen
  \bibfield  {author} {\bibinfo {author} {\bibfnamefont {I.}~\bibnamefont
  {Moreels}}, \bibinfo {author} {\bibfnamefont {K.}~\bibnamefont {Lambert}},
  \bibinfo {author} {\bibfnamefont {D.}~\bibnamefont {Smeets}}, \bibinfo
  {author} {\bibfnamefont {D.}~\bibnamefont {De~Muynck}}, \bibinfo {author}
  {\bibfnamefont {T.}~\bibnamefont {Nollet}}, \bibinfo {author} {\bibfnamefont
  {J.~C.}\ \bibnamefont {Martins}}, \bibinfo {author} {\bibfnamefont
  {F.}~\bibnamefont {Vanhaecke}}, \bibinfo {author} {\bibfnamefont
  {A.}~\bibnamefont {Vantomme}}, \bibinfo {author} {\bibfnamefont
  {C.}~\bibnamefont {Delerue}}, \bibinfo {author} {\bibfnamefont
  {G.}~\bibnamefont {Allan}},\ and\ \bibinfo {author} {\bibfnamefont
  {Z.}~\bibnamefont {Hens}},\ }\bibfield  {title} {\bibinfo {title}
  {Size-{{Dependent Optical Properties}} of {{Colloidal PbS Quantum Dots}}},\
  }\href {https://doi.org/10.1021/nn900863a} {\bibfield  {journal} {\bibinfo
  {journal} {ACS Nano}\ }\textbf {\bibinfo {volume} {3}},\ \bibinfo {pages}
  {3023} (\bibinfo {year} {2009})}\BibitemShut {NoStop}%
\bibitem [{\citenamefont {Rider}\ \emph {et~al.}(2020)\citenamefont {Rider},
  \citenamefont {Sokolikova}, \citenamefont {Hanham}, \citenamefont
  {Navarro-Cía}, \citenamefont {Haynes}, \citenamefont {Lee}, \citenamefont
  {Daniele}, \citenamefont {Cestelli~Guidi}, \citenamefont {Mattevi},
  \citenamefont {Lupi},\ and\ \citenamefont {Giannini}}]{D0NR06523D}%
  \BibitemOpen
  \bibfield  {author} {\bibinfo {author} {\bibfnamefont {M.~S.}\ \bibnamefont
  {Rider}}, \bibinfo {author} {\bibfnamefont {M.}~\bibnamefont {Sokolikova}},
  \bibinfo {author} {\bibfnamefont {S.~M.}\ \bibnamefont {Hanham}}, \bibinfo
  {author} {\bibfnamefont {M.}~\bibnamefont {Navarro-Cía}}, \bibinfo {author}
  {\bibfnamefont {P.~D.}\ \bibnamefont {Haynes}}, \bibinfo {author}
  {\bibfnamefont {D.~K.~K.}\ \bibnamefont {Lee}}, \bibinfo {author}
  {\bibfnamefont {M.}~\bibnamefont {Daniele}}, \bibinfo {author} {\bibfnamefont
  {M.}~\bibnamefont {Cestelli~Guidi}}, \bibinfo {author} {\bibfnamefont
  {C.}~\bibnamefont {Mattevi}}, \bibinfo {author} {\bibfnamefont
  {S.}~\bibnamefont {Lupi}},\ and\ \bibinfo {author} {\bibfnamefont
  {V.}~\bibnamefont {Giannini}},\ }\bibfield  {title} {\bibinfo {title}
  {Experimental signature of a topological quantum dot},\ }\href
  {https://doi.org/10.1039/D0NR06523D} {\bibfield  {journal} {\bibinfo
  {journal} {Nanoscale}\ }\textbf {\bibinfo {volume} {12}},\ \bibinfo {pages}
  {22817} (\bibinfo {year} {2020})}\BibitemShut {NoStop}%
\bibitem [{\citenamefont {Rider}\ and\ \citenamefont
  {Giannini}(2021)}]{rider2021proposal}%
  \BibitemOpen
  \bibfield  {author} {\bibinfo {author} {\bibfnamefont {M.~S.}\ \bibnamefont
  {Rider}}\ and\ \bibinfo {author} {\bibfnamefont {V.}~\bibnamefont
  {Giannini}},\ }\bibfield  {title} {\bibinfo {title} {Proposal for {{THz}}
  lasing from a topological quantum dot},\ }\href
  {https://doi.org/10.1515/nanoph-2021-0292} {\bibfield  {journal} {\bibinfo
  {journal} {Nanophotonics}\ }\textbf {\bibinfo {volume} {10}},\ \bibinfo
  {pages} {3497} (\bibinfo {year} {2021})}\BibitemShut {NoStop}%
\bibitem [{\citenamefont {Zhang}\ \emph
  {et~al.}(2023{\natexlab{b}})\citenamefont {Zhang}, \citenamefont {Wu},
  \citenamefont {Liu}, \citenamefont {Yu}, \citenamefont {Yang},\ and\
  \citenamefont {Yao}}]{zhang2023encyclopedia}%
  \BibitemOpen
  \bibfield  {author} {\bibinfo {author} {\bibfnamefont {Z.}~\bibnamefont
  {Zhang}}, \bibinfo {author} {\bibfnamefont {W.}~\bibnamefont {Wu}}, \bibinfo
  {author} {\bibfnamefont {G.-B.}\ \bibnamefont {Liu}}, \bibinfo {author}
  {\bibfnamefont {Z.-M.}\ \bibnamefont {Yu}}, \bibinfo {author} {\bibfnamefont
  {S.~A.}\ \bibnamefont {Yang}},\ and\ \bibinfo {author} {\bibfnamefont
  {Y.}~\bibnamefont {Yao}},\ }\bibfield  {title} {\bibinfo {title}
  {Encyclopedia of emergent particles in 528 magnetic layer groups and 394
  magnetic rod groups},\ }\href {https://doi.org/10.1103/PhysRevB.107.075405}
  {\bibfield  {journal} {\bibinfo  {journal} {Physical Review B}\ }\textbf
  {\bibinfo {volume} {107}},\ \bibinfo {pages} {075405} (\bibinfo {year}
  {2023}{\natexlab{b}})}\BibitemShut {NoStop}%
\bibitem [{han()}]{hansupplemental}%
  \BibitemOpen
  \href@noop {} {\ }\bibinfo {note} {See Supplemental Material for calculation
  methods, symmetry analysis, robustness of corner states, and detailed
  analysis of the topological properties of TiSiCO-family monolayer, which
  inclueds
  Refs.\cite{kresse1996efficient,blochl1994projector,perdew1996generalized,perdew1998perdew,monkhorst1976speciale,
  yu2020valleylayer,heyd2003hybrid,mostofi2014updated,wu2018wanniertools,gaoIrvspObtainIrreducible2021,ZhangENPRB-2023,Bilbaoweb,
  jia2009controlled,rizzo2018topological,hu2023identifying}}\BibitemShut
  {NoStop}%
\bibitem [{\citenamefont {Ahn}\ \emph {et~al.}(2018)\citenamefont {Ahn},
  \citenamefont {Kim}, \citenamefont {Kim},\ and\ \citenamefont
  {Yang}}]{ahn2018band}%
  \BibitemOpen
  \bibfield  {author} {\bibinfo {author} {\bibfnamefont {J.}~\bibnamefont
  {Ahn}}, \bibinfo {author} {\bibfnamefont {D.}~\bibnamefont {Kim}}, \bibinfo
  {author} {\bibfnamefont {Y.}~\bibnamefont {Kim}},\ and\ \bibinfo {author}
  {\bibfnamefont {B.-J.}\ \bibnamefont {Yang}},\ }\bibfield  {title} {\bibinfo
  {title} {Band {{Topology}} and {{Linking Structure}} of {{Nodal Line
  Semimetals}} with {{Z}}${}_2$ {{Monopole Charges}}},\ }\href
  {https://doi.org/10.1103/PhysRevLett.121.106403} {\bibfield  {journal}
  {\bibinfo  {journal} {Physical Review Letters}\ }\textbf {\bibinfo {volume}
  {121}},\ \bibinfo {pages} {106403} (\bibinfo {year} {2018})}\BibitemShut
  {NoStop}%
\bibitem [{\citenamefont {Zhao}\ and\ \citenamefont
  {Lu}(2017)}]{zhaoSymmetricRealDirac2017}%
  \BibitemOpen
  \bibfield  {author} {\bibinfo {author} {\bibfnamefont {Y.~X.}\ \bibnamefont
  {Zhao}}\ and\ \bibinfo {author} {\bibfnamefont {Y.}~\bibnamefont {Lu}},\
  }\bibfield  {title} {\bibinfo {title} {P {{T}} -{{Symmetric Real Dirac
  Fermions}} and {{Semimetals}}},\ }\href
  {https://doi.org/10.1103/PhysRevLett.118.056401} {\bibfield  {journal}
  {\bibinfo  {journal} {Physical Review Letters}\ }\textbf {\bibinfo {volume}
  {118}},\ \bibinfo {pages} {056401} (\bibinfo {year} {2017})}\BibitemShut
  {NoStop}%
\bibitem [{\citenamefont {Zhu}\ \emph {et~al.}(2022)\citenamefont {Zhu},
  \citenamefont {Wu}, \citenamefont {Zhao}, \citenamefont {Chen}, \citenamefont
  {Wang}, \citenamefont {Sheng}, \citenamefont {Zhang}, \citenamefont {Zhao},\
  and\ \citenamefont {Yang}}]{zhuPhononicRealChern2022a}%
  \BibitemOpen
  \bibfield  {author} {\bibinfo {author} {\bibfnamefont {J.}~\bibnamefont
  {Zhu}}, \bibinfo {author} {\bibfnamefont {W.}~\bibnamefont {Wu}}, \bibinfo
  {author} {\bibfnamefont {J.}~\bibnamefont {Zhao}}, \bibinfo {author}
  {\bibfnamefont {C.}~\bibnamefont {Chen}}, \bibinfo {author} {\bibfnamefont
  {Q.}~\bibnamefont {Wang}}, \bibinfo {author} {\bibfnamefont {X.-L.}\
  \bibnamefont {Sheng}}, \bibinfo {author} {\bibfnamefont {L.}~\bibnamefont
  {Zhang}}, \bibinfo {author} {\bibfnamefont {Y.~X.}\ \bibnamefont {Zhao}},\
  and\ \bibinfo {author} {\bibfnamefont {S.~A.}\ \bibnamefont {Yang}},\
  }\bibfield  {title} {\bibinfo {title} {Phononic real {{Chern}} insulator with
  protected corner modes in graphynes},\ }\href
  {https://doi.org/10.1103/PhysRevB.105.085123} {\bibfield  {journal} {\bibinfo
   {journal} {Physical Review B}\ }\textbf {\bibinfo {volume} {105}},\ \bibinfo
  {pages} {085123} (\bibinfo {year} {2022})}\BibitemShut {NoStop}%
\bibitem [{\citenamefont {Hu}\ \emph {et~al.}(2023)\citenamefont {Hu},
  \citenamefont {Zhong}, \citenamefont {Zhang}, \citenamefont {Wang},\ and\
  \citenamefont {Wang}}]{hu2023identifying}%
  \BibitemOpen
  \bibfield  {author} {\bibinfo {author} {\bibfnamefont {T.}~\bibnamefont
  {Hu}}, \bibinfo {author} {\bibfnamefont {W.}~\bibnamefont {Zhong}}, \bibinfo
  {author} {\bibfnamefont {T.}~\bibnamefont {Zhang}}, \bibinfo {author}
  {\bibfnamefont {W.}~\bibnamefont {Wang}},\ and\ \bibinfo {author}
  {\bibfnamefont {Z.~F.}\ \bibnamefont {Wang}},\ }\bibfield  {title} {\bibinfo
  {title} {Identifying topological corner states in two-dimensional
  metal-organic frameworks},\ }\href
  {https://doi.org/10.1038/s41467-023-42884-1} {\bibfield  {journal} {\bibinfo
  {journal} {Nature Communications}\ }\textbf {\bibinfo {volume} {14}},\
  \bibinfo {pages} {7092} (\bibinfo {year} {2023})}\BibitemShut {NoStop}%
\bibitem [{\citenamefont {Zhang}\ \emph {et~al.}(2009)\citenamefont {Zhang},
  \citenamefont {Tang}, \citenamefont {Girit}, \citenamefont {Hao},
  \citenamefont {Martin}, \citenamefont {Zettl}, \citenamefont {Crommie},
  \citenamefont {Shen},\ and\ \citenamefont {Wang}}]{zhang2009direct}%
  \BibitemOpen
  \bibfield  {author} {\bibinfo {author} {\bibfnamefont {Y.}~\bibnamefont
  {Zhang}}, \bibinfo {author} {\bibfnamefont {T.-T.}\ \bibnamefont {Tang}},
  \bibinfo {author} {\bibfnamefont {C.}~\bibnamefont {Girit}}, \bibinfo
  {author} {\bibfnamefont {Z.}~\bibnamefont {Hao}}, \bibinfo {author}
  {\bibfnamefont {M.~C.}\ \bibnamefont {Martin}}, \bibinfo {author}
  {\bibfnamefont {A.}~\bibnamefont {Zettl}}, \bibinfo {author} {\bibfnamefont
  {M.~F.}\ \bibnamefont {Crommie}}, \bibinfo {author} {\bibfnamefont {Y.~R.}\
  \bibnamefont {Shen}},\ and\ \bibinfo {author} {\bibfnamefont
  {F.}~\bibnamefont {Wang}},\ }\bibfield  {title} {\bibinfo {title} {Direct
  observation of a widely tunable bandgap in bilayer graphene},\ }\href
  {https://doi.org/10.1038/nature08105} {\bibfield  {journal} {\bibinfo
  {journal} {Nature}\ }\textbf {\bibinfo {volume} {459}},\ \bibinfo {pages}
  {820} (\bibinfo {year} {2009})}\BibitemShut {NoStop}%
\bibitem [{\citenamefont {Bampoulis}\ \emph {et~al.}(2023)\citenamefont
  {Bampoulis}, \citenamefont {Castenmiller}, \citenamefont {Klaassen},
  \citenamefont {Van~Mil}, \citenamefont {Liu}, \citenamefont {Liu},
  \citenamefont {Yao}, \citenamefont {Ezawa}, \citenamefont {Rudenko},\ and\
  \citenamefont {Zandvliet}}]{bampoulis2023quantum}%
  \BibitemOpen
  \bibfield  {author} {\bibinfo {author} {\bibfnamefont {P.}~\bibnamefont
  {Bampoulis}}, \bibinfo {author} {\bibfnamefont {C.}~\bibnamefont
  {Castenmiller}}, \bibinfo {author} {\bibfnamefont {D.~J.}\ \bibnamefont
  {Klaassen}}, \bibinfo {author} {\bibfnamefont {J.}~\bibnamefont {Van~Mil}},
  \bibinfo {author} {\bibfnamefont {Y.}~\bibnamefont {Liu}}, \bibinfo {author}
  {\bibfnamefont {C.-C.}\ \bibnamefont {Liu}}, \bibinfo {author} {\bibfnamefont
  {Y.}~\bibnamefont {Yao}}, \bibinfo {author} {\bibfnamefont {M.}~\bibnamefont
  {Ezawa}}, \bibinfo {author} {\bibfnamefont {A.~N.}\ \bibnamefont {Rudenko}},\
  and\ \bibinfo {author} {\bibfnamefont {H.~J.~W.}\ \bibnamefont {Zandvliet}},\
  }\bibfield  {title} {\bibinfo {title} {Quantum {{Spin Hall States}} and
  {{Topological Phase Transition}} in {{Germanene}}},\ }\href
  {https://doi.org/10.1103/PhysRevLett.130.196401} {\bibfield  {journal}
  {\bibinfo  {journal} {Physical Review Letters}\ }\textbf {\bibinfo {volume}
  {130}},\ \bibinfo {pages} {196401} (\bibinfo {year} {2023})}\BibitemShut
  {NoStop}%
\bibitem [{\citenamefont {Perinetti}(2011)}]{perinetti2011optical}%
  \BibitemOpen
  \bibfield  {author} {\bibinfo {author} {\bibfnamefont {U.}~\bibnamefont
  {Perinetti}},\ }\bibfield  {title} {\bibinfo {title} {Optical {{Properties}}
  of {{Semiconductor Quantum Dots}}},\ }\href@noop {} {\  (\bibinfo {year}
  {2011})}\BibitemShut {NoStop}%
\bibitem [{\citenamefont {Shi}\ \emph {et~al.}(2022)\citenamefont {Shi},
  \citenamefont {Yoo}, \citenamefont {{Vidal-Codina}}, \citenamefont {Baik},
  \citenamefont {Cho}, \citenamefont {Nguyen}, \citenamefont {Utzat},
  \citenamefont {Han}, \citenamefont {Lindenberg}, \citenamefont {Bulovi{\'c}},
  \citenamefont {Bawendi}, \citenamefont {Peraire}, \citenamefont {Oh},\ and\
  \citenamefont {Nelson}}]{shi2022roomtemperature}%
  \BibitemOpen
  \bibfield  {author} {\bibinfo {author} {\bibfnamefont {J.}~\bibnamefont
  {Shi}}, \bibinfo {author} {\bibfnamefont {D.}~\bibnamefont {Yoo}}, \bibinfo
  {author} {\bibfnamefont {F.}~\bibnamefont {{Vidal-Codina}}}, \bibinfo
  {author} {\bibfnamefont {C.-W.}\ \bibnamefont {Baik}}, \bibinfo {author}
  {\bibfnamefont {K.-S.}\ \bibnamefont {Cho}}, \bibinfo {author} {\bibfnamefont
  {N.-C.}\ \bibnamefont {Nguyen}}, \bibinfo {author} {\bibfnamefont
  {H.}~\bibnamefont {Utzat}}, \bibinfo {author} {\bibfnamefont
  {J.}~\bibnamefont {Han}}, \bibinfo {author} {\bibfnamefont {A.~M.}\
  \bibnamefont {Lindenberg}}, \bibinfo {author} {\bibfnamefont
  {V.}~\bibnamefont {Bulovi{\'c}}}, \bibinfo {author} {\bibfnamefont {M.~G.}\
  \bibnamefont {Bawendi}}, \bibinfo {author} {\bibfnamefont {J.}~\bibnamefont
  {Peraire}}, \bibinfo {author} {\bibfnamefont {S.-H.}\ \bibnamefont {Oh}},\
  and\ \bibinfo {author} {\bibfnamefont {K.~A.}\ \bibnamefont {Nelson}},\
  }\bibfield  {title} {\bibinfo {title} {A room-temperature
  polarization-sensitive {{CMOS}} terahertz camera based on
  quantum-dot-enhanced terahertz-to-visible photon upconversion},\ }\href
  {https://doi.org/10.1038/s41565-022-01243-9} {\bibfield  {journal} {\bibinfo
  {journal} {Nature Nanotechnology}\ }\textbf {\bibinfo {volume} {17}},\
  \bibinfo {pages} {1288} (\bibinfo {year} {2022})}\BibitemShut {NoStop}%
\bibitem [{\citenamefont {Kresse}\ and\ \citenamefont
  {Furthm{\"u}ller}(1996)}]{kresse1996efficient}%
  \BibitemOpen
  \bibfield  {author} {\bibinfo {author} {\bibfnamefont {G.}~\bibnamefont
  {Kresse}}\ and\ \bibinfo {author} {\bibfnamefont {J.}~\bibnamefont
  {Furthm{\"u}ller}},\ }\bibfield  {title} {\bibinfo {title} {Efficient
  iterative schemes for ab initio total-energy calculations using a plane-wave
  basis set},\ }\href {https://doi.org/10.1103/PhysRevB.54.11169} {\bibfield
  {journal} {\bibinfo  {journal} {Physical Review B}\ }\textbf {\bibinfo
  {volume} {54}},\ \bibinfo {pages} {11169} (\bibinfo {year}
  {1996})}\BibitemShut {NoStop}%
\bibitem [{\citenamefont {Bl{\"o}chl}(1994)}]{blochl1994projector}%
  \BibitemOpen
  \bibfield  {author} {\bibinfo {author} {\bibfnamefont {P.~E.}\ \bibnamefont
  {Bl{\"o}chl}},\ }\bibfield  {title} {\bibinfo {title} {Projector
  augmented-wave method},\ }\href {https://doi.org/10.1103/PhysRevB.50.17953}
  {\bibfield  {journal} {\bibinfo  {journal} {Physical Review B}\ }\textbf
  {\bibinfo {volume} {50}},\ \bibinfo {pages} {17953} (\bibinfo {year}
  {1994})}\BibitemShut {NoStop}%
\bibitem [{\citenamefont {Perdew}\ \emph {et~al.}(1996)\citenamefont {Perdew},
  \citenamefont {Burke},\ and\ \citenamefont
  {Ernzerhof}}]{perdew1996generalized}%
  \BibitemOpen
  \bibfield  {author} {\bibinfo {author} {\bibfnamefont {J.~P.}\ \bibnamefont
  {Perdew}}, \bibinfo {author} {\bibfnamefont {K.}~\bibnamefont {Burke}},\ and\
  \bibinfo {author} {\bibfnamefont {M.}~\bibnamefont {Ernzerhof}},\ }\bibfield
  {title} {\bibinfo {title} {Generalized {{Gradient Approximation Made
  Simple}}},\ }\href {https://doi.org/10.1103/PhysRevLett.77.3865} {\bibfield
  {journal} {\bibinfo  {journal} {Physical Review Letters}\ }\textbf {\bibinfo
  {volume} {77}},\ \bibinfo {pages} {3865} (\bibinfo {year}
  {1996})}\BibitemShut {NoStop}%
\bibitem [{\citenamefont {Perdew}\ \emph {et~al.}(1998)\citenamefont {Perdew},
  \citenamefont {Burke},\ and\ \citenamefont {Ernzerhof}}]{perdew1998perdew}%
  \BibitemOpen
  \bibfield  {author} {\bibinfo {author} {\bibfnamefont {J.~P.}\ \bibnamefont
  {Perdew}}, \bibinfo {author} {\bibfnamefont {K.}~\bibnamefont {Burke}},\ and\
  \bibinfo {author} {\bibfnamefont {M.}~\bibnamefont {Ernzerhof}},\ }\bibfield
  {title} {\bibinfo {title} {Perdew, {{Burke}}, and {{Ernzerhof Reply}}:},\
  }\href {https://doi.org/10.1103/PhysRevLett.80.891} {\bibfield  {journal}
  {\bibinfo  {journal} {Physical Review Letters}\ }\textbf {\bibinfo {volume}
  {80}},\ \bibinfo {pages} {891} (\bibinfo {year} {1998})}\BibitemShut
  {NoStop}%
\bibitem [{\citenamefont {Monkhorst}\ and\ \citenamefont
  {Pack}(1976)}]{monkhorst1976speciale}%
  \BibitemOpen
  \bibfield  {author} {\bibinfo {author} {\bibfnamefont {H.~J.}\ \bibnamefont
  {Monkhorst}}\ and\ \bibinfo {author} {\bibfnamefont {J.~D.}\ \bibnamefont
  {Pack}},\ }\bibfield  {title} {\bibinfo {title} {Special points for
  {{Brillouin-zone}} integrations},\ }\href
  {https://doi.org/10.1103/PhysRevB.13.5188} {\bibfield  {journal} {\bibinfo
  {journal} {Physical Review B}\ }\textbf {\bibinfo {volume} {13}},\ \bibinfo
  {pages} {5188} (\bibinfo {year} {1976})}\BibitemShut {NoStop}%
\bibitem [{\citenamefont {Heyd}\ \emph {et~al.}(2003)\citenamefont {Heyd},
  \citenamefont {Scuseria},\ and\ \citenamefont {Ernzerhof}}]{heyd2003hybrid}%
  \BibitemOpen
  \bibfield  {author} {\bibinfo {author} {\bibfnamefont {J.}~\bibnamefont
  {Heyd}}, \bibinfo {author} {\bibfnamefont {G.~E.}\ \bibnamefont {Scuseria}},\
  and\ \bibinfo {author} {\bibfnamefont {M.}~\bibnamefont {Ernzerhof}},\
  }\bibfield  {title} {\bibinfo {title} {Hybrid functionals based on a screened
  {Coulomb} potential},\ }\href {https://doi.org/10.1063/1.1564060} {\bibfield
  {journal} {\bibinfo  {journal} {The Journal of Chemical Physics}\ }\textbf
  {\bibinfo {volume} {118}},\ \bibinfo {pages} {8207} (\bibinfo {year}
  {2003})}\BibitemShut {NoStop}%
\bibitem [{\citenamefont {Mostofi}\ \emph {et~al.}(2014)\citenamefont
  {Mostofi}, \citenamefont {Yates}, \citenamefont {Pizzi}, \citenamefont {Lee},
  \citenamefont {Souza}, \citenamefont {Vanderbilt},\ and\ \citenamefont
  {Marzari}}]{mostofi2014updated}%
  \BibitemOpen
  \bibfield  {author} {\bibinfo {author} {\bibfnamefont {A.~A.}\ \bibnamefont
  {Mostofi}}, \bibinfo {author} {\bibfnamefont {J.~R.}\ \bibnamefont {Yates}},
  \bibinfo {author} {\bibfnamefont {G.}~\bibnamefont {Pizzi}}, \bibinfo
  {author} {\bibfnamefont {Y.-S.}\ \bibnamefont {Lee}}, \bibinfo {author}
  {\bibfnamefont {I.}~\bibnamefont {Souza}}, \bibinfo {author} {\bibfnamefont
  {D.}~\bibnamefont {Vanderbilt}},\ and\ \bibinfo {author} {\bibfnamefont
  {N.}~\bibnamefont {Marzari}},\ }\bibfield  {title} {\bibinfo {title} {An
  updated version of wannier90: {{A}} tool for obtaining maximally-localised
  {{Wannier}} functions},\ }\href {https://doi.org/10.1016/j.cpc.2014.05.003}
  {\bibfield  {journal} {\bibinfo  {journal} {Computer Physics Communications}\
  }\textbf {\bibinfo {volume} {185}},\ \bibinfo {pages} {2309} (\bibinfo {year}
  {2014})}\BibitemShut {NoStop}%
\bibitem [{\citenamefont {Wu}\ \emph {et~al.}(2018)\citenamefont {Wu},
  \citenamefont {Zhang}, \citenamefont {Song}, \citenamefont {Troyer},\ and\
  \citenamefont {Soluyanov}}]{wu2018wanniertools}%
  \BibitemOpen
  \bibfield  {author} {\bibinfo {author} {\bibfnamefont {Q.}~\bibnamefont
  {Wu}}, \bibinfo {author} {\bibfnamefont {S.}~\bibnamefont {Zhang}}, \bibinfo
  {author} {\bibfnamefont {H.-F.}\ \bibnamefont {Song}}, \bibinfo {author}
  {\bibfnamefont {M.}~\bibnamefont {Troyer}},\ and\ \bibinfo {author}
  {\bibfnamefont {A.~A.}\ \bibnamefont {Soluyanov}},\ }\bibfield  {title}
  {\bibinfo {title} {{{WannierTools}}: {{An}} open-source software package for
  novel topological materials},\ }\href
  {https://doi.org/10.1016/j.cpc.2017.09.033} {\bibfield  {journal} {\bibinfo
  {journal} {Computer Physics Communications}\ }\textbf {\bibinfo {volume}
  {224}},\ \bibinfo {pages} {405} (\bibinfo {year} {2018})}\BibitemShut
  {NoStop}%
\bibitem [{\citenamefont {Gao}\ \emph {et~al.}(2021)\citenamefont {Gao},
  \citenamefont {Wu}, \citenamefont {Persson},\ and\ \citenamefont
  {Wang}}]{gaoIrvspObtainIrreducible2021}%
  \BibitemOpen
  \bibfield  {author} {\bibinfo {author} {\bibfnamefont {J.}~\bibnamefont
  {Gao}}, \bibinfo {author} {\bibfnamefont {Q.}~\bibnamefont {Wu}}, \bibinfo
  {author} {\bibfnamefont {C.}~\bibnamefont {Persson}},\ and\ \bibinfo {author}
  {\bibfnamefont {Z.}~\bibnamefont {Wang}},\ }\bibfield  {title} {\bibinfo
  {title} {Irvsp: {{To}} obtain irreducible representations of electronic
  states in the {{VASP}}},\ }\href {https://doi.org/10.1016/j.cpc.2020.107760}
  {\bibfield  {journal} {\bibinfo  {journal} {Computer Physics Communications}\
  }\textbf {\bibinfo {volume} {261}},\ \bibinfo {pages} {107760} (\bibinfo
  {year} {2021})}\BibitemShut {NoStop}%
\bibitem [{\citenamefont {Zhang}\ \emph
  {et~al.}(2023{\natexlab{c}})\citenamefont {Zhang}, \citenamefont {Wu},
  \citenamefont {Liu}, \citenamefont {Yu}, \citenamefont {Yang},\ and\
  \citenamefont {Yao}}]{ZhangENPRB-2023}%
  \BibitemOpen
  \bibfield  {author} {\bibinfo {author} {\bibfnamefont {Z.}~\bibnamefont
  {Zhang}}, \bibinfo {author} {\bibfnamefont {W.}~\bibnamefont {Wu}}, \bibinfo
  {author} {\bibfnamefont {G.-B.}\ \bibnamefont {Liu}}, \bibinfo {author}
  {\bibfnamefont {Z.-M.}\ \bibnamefont {Yu}}, \bibinfo {author} {\bibfnamefont
  {S.~A.}\ \bibnamefont {Yang}},\ and\ \bibinfo {author} {\bibfnamefont
  {Y.}~\bibnamefont {Yao}},\ }\bibfield  {title} {\bibinfo {title}
  {Encyclopedia of emergent particles in 528 magnetic layer groups and 394
  magnetic rod groups},\ }\href {https://doi.org/10.1103/PhysRevB.107.075405}
  {\bibfield  {journal} {\bibinfo  {journal} {Phys. Rev. B}\ }\textbf {\bibinfo
  {volume} {107}},\ \bibinfo {pages} {075405} (\bibinfo {year}
  {2023}{\natexlab{c}})}\BibitemShut {NoStop}%
\bibitem [{Bil()}]{Bilbaoweb}%
  \BibitemOpen
  \href@noop {} {\ }\bibinfo {note} {Bilbao Crystallographic Server, under
  ``Subperiodic Groups: Layer, Rod and Frieze Groups"}\BibitemShut {NoStop}%
\bibitem [{\citenamefont {Jia}\ \emph {et~al.}(2009)\citenamefont {Jia},
  \citenamefont {Hofmann}, \citenamefont {Meunier}, \citenamefont {Sumpter},
  \citenamefont {{Campos-Delgado}}, \citenamefont {{Romo-Herrera}},
  \citenamefont {Son}, \citenamefont {Hsieh}, \citenamefont {Reina},
  \citenamefont {Kong}, \citenamefont {Terrones},\ and\ \citenamefont
  {Dresselhaus}}]{jia2009controlled}%
  \BibitemOpen
  \bibfield  {author} {\bibinfo {author} {\bibfnamefont {X.}~\bibnamefont
  {Jia}}, \bibinfo {author} {\bibfnamefont {M.}~\bibnamefont {Hofmann}},
  \bibinfo {author} {\bibfnamefont {V.}~\bibnamefont {Meunier}}, \bibinfo
  {author} {\bibfnamefont {B.~G.}\ \bibnamefont {Sumpter}}, \bibinfo {author}
  {\bibfnamefont {J.}~\bibnamefont {{Campos-Delgado}}}, \bibinfo {author}
  {\bibfnamefont {J.~M.}\ \bibnamefont {{Romo-Herrera}}}, \bibinfo {author}
  {\bibfnamefont {H.}~\bibnamefont {Son}}, \bibinfo {author} {\bibfnamefont
  {Y.-P.}\ \bibnamefont {Hsieh}}, \bibinfo {author} {\bibfnamefont
  {A.}~\bibnamefont {Reina}}, \bibinfo {author} {\bibfnamefont
  {J.}~\bibnamefont {Kong}}, \bibinfo {author} {\bibfnamefont {M.}~\bibnamefont
  {Terrones}},\ and\ \bibinfo {author} {\bibfnamefont {M.~S.}\ \bibnamefont
  {Dresselhaus}},\ }\bibfield  {title} {\bibinfo {title} {Controlled
  {{Formation}} of {{Sharp Zigzag}} and {{Armchair Edges}} in {{Graphitic
  Nanoribbons}}},\ }\href {https://doi.org/10.1126/science.1166862} {\bibfield
  {journal} {\bibinfo  {journal} {Science}\ }\textbf {\bibinfo {volume}
  {323}},\ \bibinfo {pages} {1701} (\bibinfo {year} {2009})}\BibitemShut
  {NoStop}%
\bibitem [{\citenamefont {Rizzo}\ \emph {et~al.}(2018)\citenamefont {Rizzo},
  \citenamefont {Veber}, \citenamefont {Cao}, \citenamefont {Bronner},
  \citenamefont {Chen}, \citenamefont {Zhao}, \citenamefont {Rodriguez},
  \citenamefont {Louie}, \citenamefont {Crommie},\ and\ \citenamefont
  {Fischer}}]{rizzo2018topological}%
  \BibitemOpen
  \bibfield  {author} {\bibinfo {author} {\bibfnamefont {D.~J.}\ \bibnamefont
  {Rizzo}}, \bibinfo {author} {\bibfnamefont {G.}~\bibnamefont {Veber}},
  \bibinfo {author} {\bibfnamefont {T.}~\bibnamefont {Cao}}, \bibinfo {author}
  {\bibfnamefont {C.}~\bibnamefont {Bronner}}, \bibinfo {author} {\bibfnamefont
  {T.}~\bibnamefont {Chen}}, \bibinfo {author} {\bibfnamefont {F.}~\bibnamefont
  {Zhao}}, \bibinfo {author} {\bibfnamefont {H.}~\bibnamefont {Rodriguez}},
  \bibinfo {author} {\bibfnamefont {S.~G.}\ \bibnamefont {Louie}}, \bibinfo
  {author} {\bibfnamefont {M.~F.}\ \bibnamefont {Crommie}},\ and\ \bibinfo
  {author} {\bibfnamefont {F.~R.}\ \bibnamefont {Fischer}},\ }\bibfield
  {title} {\bibinfo {title} {Topological band engineering of graphene
  nanoribbons},\ }\href {https://doi.org/10.1038/s41586-018-0376-8} {\bibfield
  {journal} {\bibinfo  {journal} {Nature}\ }\textbf {\bibinfo {volume} {560}},\
  \bibinfo {pages} {204} (\bibinfo {year} {2018})}\BibitemShut {NoStop}%
\end{thebibliography}%


\begin{thebibliography}{15}%
\makeatletter
\providecommand \@ifxundefined [1]{%
 \@ifx{#1\undefined}
}%
\providecommand \@ifnum [1]{%
 \ifnum #1\expandafter \@firstoftwo
 \else \expandafter \@secondoftwo
 \fi
}%
\providecommand \@ifx [1]{%
 \ifx #1\expandafter \@firstoftwo
 \else \expandafter \@secondoftwo
 \fi
}%
\providecommand \natexlab [1]{#1}%
\providecommand \enquote  [1]{``#1''}%
\providecommand \bibnamefont  [1]{#1}%
\providecommand \bibfnamefont [1]{#1}%
\providecommand \citenamefont [1]{#1}%
\providecommand \href@noop [0]{\@secondoftwo}%
\providecommand \href [0]{\begingroup \@sanitize@url \@href}%
\providecommand \@href[1]{\@@startlink{#1}\@@href}%
\providecommand \@@href[1]{\endgroup#1\@@endlink}%
\providecommand \@sanitize@url [0]{\catcode `\\12\catcode `\$12\catcode
  `\&12\catcode `\#12\catcode `\^12\catcode `\_12\catcode `\%12\relax}%
\providecommand \@@startlink[1]{}%
\providecommand \@@endlink[0]{}%
\providecommand \url  [0]{\begingroup\@sanitize@url \@url }%
\providecommand \@url [1]{\endgroup\@href {#1}{\urlprefix }}%
\providecommand \urlprefix  [0]{URL }%
\providecommand \Eprint [0]{\href }%
\providecommand \doibase [0]{https://doi.org/}%
\providecommand \selectlanguage [0]{\@gobble}%
\providecommand \bibinfo  [0]{\@secondoftwo}%
\providecommand \bibfield  [0]{\@secondoftwo}%
\providecommand \translation [1]{[#1]}%
\providecommand \BibitemOpen [0]{}%
\providecommand \bibitemStop [0]{}%
\providecommand \bibitemNoStop [0]{.\EOS\space}%
\providecommand \EOS [0]{\spacefactor3000\relax}%
\providecommand \BibitemShut  [1]{\csname bibitem#1\endcsname}%
\let\auto@bib@innerbib\@empty
\bibitem [{\citenamefont {Kresse}\ and\ \citenamefont
  {Furthm{\"u}ller}(1996)}]{kresse1996efficient}%
  \BibitemOpen
  \bibfield  {author} {\bibinfo {author} {\bibfnamefont {G.}~\bibnamefont
  {Kresse}}\ and\ \bibinfo {author} {\bibfnamefont {J.}~\bibnamefont
  {Furthm{\"u}ller}},\ }\bibfield  {title} {\bibinfo {title} {Efficient
  iterative schemes for ab initio total-energy calculations using a plane-wave
  basis set},\ }\href {https://doi.org/10.1103/PhysRevB.54.11169} {\bibfield
  {journal} {\bibinfo  {journal} {Physical Review B}\ }\textbf {\bibinfo
  {volume} {54}},\ \bibinfo {pages} {11169} (\bibinfo {year}
  {1996})}\BibitemShut {NoStop}%
\bibitem [{\citenamefont {Bl{\"o}chl}(1994)}]{blochl1994projector}%
  \BibitemOpen
  \bibfield  {author} {\bibinfo {author} {\bibfnamefont {P.~E.}\ \bibnamefont
  {Bl{\"o}chl}},\ }\bibfield  {title} {\bibinfo {title} {Projector
  augmented-wave method},\ }\href {https://doi.org/10.1103/PhysRevB.50.17953}
  {\bibfield  {journal} {\bibinfo  {journal} {Physical Review B}\ }\textbf
  {\bibinfo {volume} {50}},\ \bibinfo {pages} {17953} (\bibinfo {year}
  {1994})}\BibitemShut {NoStop}%
\bibitem [{\citenamefont {Perdew}\ \emph {et~al.}(1996)\citenamefont {Perdew},
  \citenamefont {Burke},\ and\ \citenamefont
  {Ernzerhof}}]{perdew1996generalized}%
  \BibitemOpen
  \bibfield  {author} {\bibinfo {author} {\bibfnamefont {J.~P.}\ \bibnamefont
  {Perdew}}, \bibinfo {author} {\bibfnamefont {K.}~\bibnamefont {Burke}},\ and\
  \bibinfo {author} {\bibfnamefont {M.}~\bibnamefont {Ernzerhof}},\ }\bibfield
  {title} {\bibinfo {title} {Generalized {{Gradient Approximation Made
  Simple}}},\ }\href {https://doi.org/10.1103/PhysRevLett.77.3865} {\bibfield
  {journal} {\bibinfo  {journal} {Physical Review Letters}\ }\textbf {\bibinfo
  {volume} {77}},\ \bibinfo {pages} {3865} (\bibinfo {year}
  {1996})}\BibitemShut {NoStop}%
\bibitem [{\citenamefont {Perdew}\ \emph {et~al.}(1998)\citenamefont {Perdew},
  \citenamefont {Burke},\ and\ \citenamefont {Ernzerhof}}]{perdew1998perdew}%
  \BibitemOpen
  \bibfield  {author} {\bibinfo {author} {\bibfnamefont {J.~P.}\ \bibnamefont
  {Perdew}}, \bibinfo {author} {\bibfnamefont {K.}~\bibnamefont {Burke}},\ and\
  \bibinfo {author} {\bibfnamefont {M.}~\bibnamefont {Ernzerhof}},\ }\bibfield
  {title} {\bibinfo {title} {Perdew, {{Burke}}, and {{Ernzerhof Reply}}:},\
  }\href {https://doi.org/10.1103/PhysRevLett.80.891} {\bibfield  {journal}
  {\bibinfo  {journal} {Physical Review Letters}\ }\textbf {\bibinfo {volume}
  {80}},\ \bibinfo {pages} {891} (\bibinfo {year} {1998})}\BibitemShut
  {NoStop}%
\bibitem [{\citenamefont {Monkhorst}\ and\ \citenamefont
  {Pack}(1976)}]{monkhorst1976speciale}%
  \BibitemOpen
  \bibfield  {author} {\bibinfo {author} {\bibfnamefont {H.~J.}\ \bibnamefont
  {Monkhorst}}\ and\ \bibinfo {author} {\bibfnamefont {J.~D.}\ \bibnamefont
  {Pack}},\ }\bibfield  {title} {\bibinfo {title} {Special points for
  {{Brillouin-zone}} integrations},\ }\href
  {https://doi.org/10.1103/PhysRevB.13.5188} {\bibfield  {journal} {\bibinfo
  {journal} {Physical Review B}\ }\textbf {\bibinfo {volume} {13}},\ \bibinfo
  {pages} {5188} (\bibinfo {year} {1976})}\BibitemShut {NoStop}%
\bibitem [{\citenamefont {Yu}\ \emph {et~al.}(2020)\citenamefont {Yu},
  \citenamefont {Guan}, \citenamefont {Sheng}, \citenamefont {Gao},\ and\
  \citenamefont {Yang}}]{yu2020valleylayer}%
  \BibitemOpen
  \bibfield  {author} {\bibinfo {author} {\bibfnamefont {Z.-M.}\ \bibnamefont
  {Yu}}, \bibinfo {author} {\bibfnamefont {S.}~\bibnamefont {Guan}}, \bibinfo
  {author} {\bibfnamefont {X.-L.}\ \bibnamefont {Sheng}}, \bibinfo {author}
  {\bibfnamefont {W.}~\bibnamefont {Gao}},\ and\ \bibinfo {author}
  {\bibfnamefont {S.~A.}\ \bibnamefont {Yang}},\ }\bibfield  {title} {\bibinfo
  {title} {Valley-{{Layer Coupling}}: {{A New Design Principle}} for
  {{Valleytronics}}},\ }\href {https://doi.org/10.1103/PhysRevLett.124.037701}
  {\bibfield  {journal} {\bibinfo  {journal} {Physical Review Letters}\
  }\textbf {\bibinfo {volume} {124}},\ \bibinfo {pages} {037701} (\bibinfo
  {year} {2020})}\BibitemShut {NoStop}%
\bibitem [{\citenamefont {Heyd}\ \emph {et~al.}(2003)\citenamefont {Heyd},
  \citenamefont {Scuseria},\ and\ \citenamefont {Ernzerhof}}]{heyd2003hybrid}%
  \BibitemOpen
  \bibfield  {author} {\bibinfo {author} {\bibfnamefont {J.}~\bibnamefont
  {Heyd}}, \bibinfo {author} {\bibfnamefont {G.~E.}\ \bibnamefont {Scuseria}},\
  and\ \bibinfo {author} {\bibfnamefont {M.}~\bibnamefont {Ernzerhof}},\
  }\bibfield  {title} {\bibinfo {title} {Hybrid functionals based on a screened
  {Coulomb} potential},\ }\href {https://doi.org/10.1063/1.1564060} {\bibfield
  {journal} {\bibinfo  {journal} {The Journal of Chemical Physics}\ }\textbf
  {\bibinfo {volume} {118}},\ \bibinfo {pages} {8207} (\bibinfo {year}
  {2003})}\BibitemShut {NoStop}%
\bibitem [{\citenamefont {Mostofi}\ \emph {et~al.}(2014)\citenamefont
  {Mostofi}, \citenamefont {Yates}, \citenamefont {Pizzi}, \citenamefont {Lee},
  \citenamefont {Souza}, \citenamefont {Vanderbilt},\ and\ \citenamefont
  {Marzari}}]{mostofi2014updated}%
  \BibitemOpen
  \bibfield  {author} {\bibinfo {author} {\bibfnamefont {A.~A.}\ \bibnamefont
  {Mostofi}}, \bibinfo {author} {\bibfnamefont {J.~R.}\ \bibnamefont {Yates}},
  \bibinfo {author} {\bibfnamefont {G.}~\bibnamefont {Pizzi}}, \bibinfo
  {author} {\bibfnamefont {Y.-S.}\ \bibnamefont {Lee}}, \bibinfo {author}
  {\bibfnamefont {I.}~\bibnamefont {Souza}}, \bibinfo {author} {\bibfnamefont
  {D.}~\bibnamefont {Vanderbilt}},\ and\ \bibinfo {author} {\bibfnamefont
  {N.}~\bibnamefont {Marzari}},\ }\bibfield  {title} {\bibinfo {title} {An
  updated version of wannier90: {{A}} tool for obtaining maximally-localised
  {{Wannier}} functions},\ }\href {https://doi.org/10.1016/j.cpc.2014.05.003}
  {\bibfield  {journal} {\bibinfo  {journal} {Computer Physics Communications}\
  }\textbf {\bibinfo {volume} {185}},\ \bibinfo {pages} {2309} (\bibinfo {year}
  {2014})}\BibitemShut {NoStop}%
\bibitem [{\citenamefont {Wu}\ \emph {et~al.}(2018)\citenamefont {Wu},
  \citenamefont {Zhang}, \citenamefont {Song}, \citenamefont {Troyer},\ and\
  \citenamefont {Soluyanov}}]{wu2018wanniertools}%
  \BibitemOpen
  \bibfield  {author} {\bibinfo {author} {\bibfnamefont {Q.}~\bibnamefont
  {Wu}}, \bibinfo {author} {\bibfnamefont {S.}~\bibnamefont {Zhang}}, \bibinfo
  {author} {\bibfnamefont {H.-F.}\ \bibnamefont {Song}}, \bibinfo {author}
  {\bibfnamefont {M.}~\bibnamefont {Troyer}},\ and\ \bibinfo {author}
  {\bibfnamefont {A.~A.}\ \bibnamefont {Soluyanov}},\ }\bibfield  {title}
  {\bibinfo {title} {{{WannierTools}}: {{An}} open-source software package for
  novel topological materials},\ }\href
  {https://doi.org/10.1016/j.cpc.2017.09.033} {\bibfield  {journal} {\bibinfo
  {journal} {Computer Physics Communications}\ }\textbf {\bibinfo {volume}
  {224}},\ \bibinfo {pages} {405} (\bibinfo {year} {2018})}\BibitemShut
  {NoStop}%
\bibitem [{\citenamefont {Gao}\ \emph {et~al.}(2021)\citenamefont {Gao},
  \citenamefont {Wu}, \citenamefont {Persson},\ and\ \citenamefont
  {Wang}}]{gaoIrvspObtainIrreducible2021}%
  \BibitemOpen
  \bibfield  {author} {\bibinfo {author} {\bibfnamefont {J.}~\bibnamefont
  {Gao}}, \bibinfo {author} {\bibfnamefont {Q.}~\bibnamefont {Wu}}, \bibinfo
  {author} {\bibfnamefont {C.}~\bibnamefont {Persson}},\ and\ \bibinfo {author}
  {\bibfnamefont {Z.}~\bibnamefont {Wang}},\ }\bibfield  {title} {\bibinfo
  {title} {Irvsp: {{To}} obtain irreducible representations of electronic
  states in the {{VASP}}},\ }\href {https://doi.org/10.1016/j.cpc.2020.107760}
  {\bibfield  {journal} {\bibinfo  {journal} {Computer Physics Communications}\
  }\textbf {\bibinfo {volume} {261}},\ \bibinfo {pages} {107760} (\bibinfo
  {year} {2021})}\BibitemShut {NoStop}%
\bibitem [{\citenamefont {Zhang}\ \emph {et~al.}(2023)\citenamefont {Zhang},
  \citenamefont {Wu}, \citenamefont {Liu}, \citenamefont {Yu}, \citenamefont
  {Yang},\ and\ \citenamefont {Yao}}]{ZhangENPRB-2023}%
  \BibitemOpen
  \bibfield  {author} {\bibinfo {author} {\bibfnamefont {Z.}~\bibnamefont
  {Zhang}}, \bibinfo {author} {\bibfnamefont {W.}~\bibnamefont {Wu}}, \bibinfo
  {author} {\bibfnamefont {G.-B.}\ \bibnamefont {Liu}}, \bibinfo {author}
  {\bibfnamefont {Z.-M.}\ \bibnamefont {Yu}}, \bibinfo {author} {\bibfnamefont
  {S.~A.}\ \bibnamefont {Yang}},\ and\ \bibinfo {author} {\bibfnamefont
  {Y.}~\bibnamefont {Yao}},\ }\bibfield  {title} {\bibinfo {title}
  {Encyclopedia of emergent particles in 528 magnetic layer groups and 394
  magnetic rod groups},\ }\href {https://doi.org/10.1103/PhysRevB.107.075405}
  {\bibfield  {journal} {\bibinfo  {journal} {Phys. Rev. B}\ }\textbf {\bibinfo
  {volume} {107}},\ \bibinfo {pages} {075405} (\bibinfo {year}
  {2023})}\BibitemShut {NoStop}%
\bibitem [{Bil()}]{Bilbaoweb}%
  \BibitemOpen
  \href@noop {} {\ }\bibinfo {note} {Bilbao Crystallographic Server, under
  ``Subperiodic Groups: Layer, Rod and Frieze Groups"}\BibitemShut {NoStop}%
\bibitem [{\citenamefont {Jia}\ \emph {et~al.}(2009)\citenamefont {Jia},
  \citenamefont {Hofmann}, \citenamefont {Meunier}, \citenamefont {Sumpter},
  \citenamefont {{Campos-Delgado}}, \citenamefont {{Romo-Herrera}},
  \citenamefont {Son}, \citenamefont {Hsieh}, \citenamefont {Reina},
  \citenamefont {Kong}, \citenamefont {Terrones},\ and\ \citenamefont
  {Dresselhaus}}]{jia2009controlled}%
  \BibitemOpen
  \bibfield  {author} {\bibinfo {author} {\bibfnamefont {X.}~\bibnamefont
  {Jia}}, \bibinfo {author} {\bibfnamefont {M.}~\bibnamefont {Hofmann}},
  \bibinfo {author} {\bibfnamefont {V.}~\bibnamefont {Meunier}}, \bibinfo
  {author} {\bibfnamefont {B.~G.}\ \bibnamefont {Sumpter}}, \bibinfo {author}
  {\bibfnamefont {J.}~\bibnamefont {{Campos-Delgado}}}, \bibinfo {author}
  {\bibfnamefont {J.~M.}\ \bibnamefont {{Romo-Herrera}}}, \bibinfo {author}
  {\bibfnamefont {H.}~\bibnamefont {Son}}, \bibinfo {author} {\bibfnamefont
  {Y.-P.}\ \bibnamefont {Hsieh}}, \bibinfo {author} {\bibfnamefont
  {A.}~\bibnamefont {Reina}}, \bibinfo {author} {\bibfnamefont
  {J.}~\bibnamefont {Kong}}, \bibinfo {author} {\bibfnamefont {M.}~\bibnamefont
  {Terrones}},\ and\ \bibinfo {author} {\bibfnamefont {M.~S.}\ \bibnamefont
  {Dresselhaus}},\ }\bibfield  {title} {\bibinfo {title} {Controlled
  {{Formation}} of {{Sharp Zigzag}} and {{Armchair Edges}} in {{Graphitic
  Nanoribbons}}},\ }\href {https://doi.org/10.1126/science.1166862} {\bibfield
  {journal} {\bibinfo  {journal} {Science}\ }\textbf {\bibinfo {volume}
  {323}},\ \bibinfo {pages} {1701} (\bibinfo {year} {2009})}\BibitemShut
  {NoStop}%
\bibitem [{\citenamefont {Rizzo}\ \emph {et~al.}(2018)\citenamefont {Rizzo},
  \citenamefont {Veber}, \citenamefont {Cao}, \citenamefont {Bronner},
  \citenamefont {Chen}, \citenamefont {Zhao}, \citenamefont {Rodriguez},
  \citenamefont {Louie}, \citenamefont {Crommie},\ and\ \citenamefont
  {Fischer}}]{rizzo2018topological}%
  \BibitemOpen
  \bibfield  {author} {\bibinfo {author} {\bibfnamefont {D.~J.}\ \bibnamefont
  {Rizzo}}, \bibinfo {author} {\bibfnamefont {G.}~\bibnamefont {Veber}},
  \bibinfo {author} {\bibfnamefont {T.}~\bibnamefont {Cao}}, \bibinfo {author}
  {\bibfnamefont {C.}~\bibnamefont {Bronner}}, \bibinfo {author} {\bibfnamefont
  {T.}~\bibnamefont {Chen}}, \bibinfo {author} {\bibfnamefont {F.}~\bibnamefont
  {Zhao}}, \bibinfo {author} {\bibfnamefont {H.}~\bibnamefont {Rodriguez}},
  \bibinfo {author} {\bibfnamefont {S.~G.}\ \bibnamefont {Louie}}, \bibinfo
  {author} {\bibfnamefont {M.~F.}\ \bibnamefont {Crommie}},\ and\ \bibinfo
  {author} {\bibfnamefont {F.~R.}\ \bibnamefont {Fischer}},\ }\bibfield
  {title} {\bibinfo {title} {Topological band engineering of graphene
  nanoribbons},\ }\href {https://doi.org/10.1038/s41586-018-0376-8} {\bibfield
  {journal} {\bibinfo  {journal} {Nature}\ }\textbf {\bibinfo {volume} {560}},\
  \bibinfo {pages} {204} (\bibinfo {year} {2018})}\BibitemShut {NoStop}%
\bibitem [{\citenamefont {Hu}\ \emph {et~al.}(2023)\citenamefont {Hu},
  \citenamefont {Zhong}, \citenamefont {Zhang}, \citenamefont {Wang},\ and\
  \citenamefont {Wang}}]{hu2023identifying}%
  \BibitemOpen
  \bibfield  {author} {\bibinfo {author} {\bibfnamefont {T.}~\bibnamefont
  {Hu}}, \bibinfo {author} {\bibfnamefont {W.}~\bibnamefont {Zhong}}, \bibinfo
  {author} {\bibfnamefont {T.}~\bibnamefont {Zhang}}, \bibinfo {author}
  {\bibfnamefont {W.}~\bibnamefont {Wang}},\ and\ \bibinfo {author}
  {\bibfnamefont {Z.~F.}\ \bibnamefont {Wang}},\ }\bibfield  {title} {\bibinfo
  {title} {Identifying topological corner states in two-dimensional
  metal-organic frameworks},\ }\href
  {https://doi.org/10.1038/s41467-023-42884-1} {\bibfield  {journal} {\bibinfo
  {journal} {Nature Communications}\ }\textbf {\bibinfo {volume} {14}},\
  \bibinfo {pages} {7092} (\bibinfo {year} {2023})}\BibitemShut {NoStop}%
\end{thebibliography}%
\end{document}